\PassOptionsToPackage{dvipsnames,svgnames,table}{xcolor}

\documentclass[a4paper,11pt]{article}

\usepackage{jheppub}
\usepackage{hyperref}
\usepackage{orcidlink}

\usepackage{lineno}
\usepackage{graphicx}
\usepackage{mathtools}
\usepackage{braket}
\usepackage{pifont}

\usepackage{tikz}
\usepackage[compat=1.1.0]{tikz-feynman}
\usepackage{contour}
\usepackage{enumitem}
\usepackage{subcaption}

\usetikzlibrary{arrows}

\newcommand{\alp}{\alpha^\prime}
\newcommand{\pa}{\partial}

\title{High-Energy Pion Scattering in Holographic QCD: A~Comparison with Experimental Data}

\author[a]{Adi Armoni
\orcidlink{0000-0002-8105-0645}}
\affiliation[a]{Department of Physics, Swansea University,\\Singleton Park, Swansea SA2 8PP, UK}
\emailAdd{a.armoni@swansea.ac.uk}

\author[b]{and Dorin Weissman\,\orcidlink{0000-0001-9697-0252}}
\affiliation[b]{INFN Sezione di Napoli,\\Monte S. Angelo, Via Cintia, 80126 Naples, Italy}
\emailAdd{dorin.weissman@na.infn.it}

\abstract{Following Polchinski and Strassler \cite{Polchinski:2001tt} and our previous work \cite{Armoni:2024gqv}, we study high-energy pion scattering in the holographic QCD hard-wall model.\,\,In particular, we focus on comparing our predictions for the angular dependence of $\pi^{+} \pi^{-} \to \pi^{+} \pi^{-}$ scattering with experimental data extracted from the process $\pi^{-} p \to \pi^{+} \pi^{-} n$.\,\,Having previously shown that our approach reproduces the constituent counting rule found in QCD, we now observe qualitative agreement between our predictions and the extracted data in the high-energy fixed-angle regime.\,\,We also provide predictions for all other $2$-to-$2$ pion scattering processes. Our approach can be extended to a broader range of meson and glueball scattering processes in various holographic QCD models.
}

\begin{document}

\maketitle
\flushbottom
\clearpage
\section{Introduction} \label{sec:1_intro}

Since the introduction of holography \cite{Maldacena:1997re}, growing evidence now suggests that string theory may provide a dual description of QCD in the low-energy regime (for pioneering results, see references \cite{Witten:1998zw, Karch:2002sh, Sakai:2004cn}; for a selection of reviews, see references \cite{Aharony:1999ti, Kim:2012ey, Ammon:2015wua}).\,\,Originally, however, string theory emerged from efforts to study high-energy meson scattering.\,\,In this context, the Veneziano formula \cite{Veneziano:1968yb} achieved partial success in the Regge regime \cite{Green:1987sp}.\,\,Going beyond the Regge regime to the high-energy fixed-angle regime, dimensional counting arguments in QCD predict a leading-order power-law scaling in the Mandelstam variable $s$ for meson scattering amplitudes $\mathcal{A}$, as captured by the constituent counting rule \cite{Matveev:1973ra, Brodsky:1973kr, Brodsky:1974vy}:
\begin{equation} \label{eq:1_CCR}
    \mathcal{A}(s, \theta_{i}) \sim s^{2 - \frac{m}{2}} f(\theta_{i}),
\end{equation}
where $m$ is the minimum number of hard constituents involved in the scattering, and $f(\theta_{i})$ encodes the dependence on the scattering angles $\theta_i$.\,\,The constituent counting rule \eqref{eq:1_CCR} appears to be consistent with all experimental studies to date when applied appropriately. For recent discussion, see reference \cite{Reed:2020dpf} and references therein.\,\,By contrast, string theory in Minkowski space demonstrates exponential rather than power-law suppression in $s$ in the high-energy fixed-angle regime.

To revisit this situation in the context of holography, Polchinski and Strassler proposed studying scalar glueball scattering by analysing the dual dilaton superstring scattering in five-dimensional AdS space with an IR cut-off \cite{Polchinski:2001tt}.\,\,Using this phenomenological approach, which models QCD's UV behaviour via a strongly coupled CFT, Polchinski and Strassler recovered the constituent counting rule.\,\,More recently, in collaboration with Sugimoto \cite{Armoni:2024gqv}, we applied their proposal to pion and $\rho$-meson scattering in a related holographic model, also recovering the constituent counting rule.\,\,However, for processes involving $\rho$-mesons, we had to generalise the original proposal \cite{Polchinski:2001tt} to reproduce the constituent counting rule.

In this paper, we compare the predictions of our model \cite{Armoni:2024gqv}, reformulated in the context of the hard-wall model, for the angular dependence of $\pi^{+} \pi^{-} \to \pi^{+} \pi^{-}$ scattering with experimental data. Since there is no direct data for $\pi^{+} \pi^{-} \to \pi^{+} \pi^{-}$ scattering, we use data which is indirectly extracted from $\pi^{-} p \to \pi^{+} \pi^{-} n$ measurements reported in reference \cite{Bromberg:1983he}.\,\,We extract the data using two different models, the simpler of which assumes only a Yukawa-type nucleon-nucleon interaction mediated by the pion (details will be provided in \hyperref[sec:3_experimental]{Section 3}).\,\,The extracted data are thus off-shell. However, we expect deviations from on-shell behaviour to be qualitatively negligible when the Mandelstam variables in the underlying pion process far exceed the pion mass \cite{Martin:1976mb, Stampke:1982cf}.

Regarding our holographic model from reference \cite{Armoni:2024gqv}, note that it is formulated within a five-dimensional asymptotically AdS (AAdS) background consistent with confinement and chiral symmetry breaking, hereafter referred to as AAdS space.\,\,Our predictions in the limit of $s\to\infty$ are based on an ansatz for tree-level superstring scattering to leading order in $\alp / R_5^2$ in this AAdS space (where $\alp$ is the Regge slope and $R_5$ is the AdS radius), and rely only on~its AdS region, which governs the leading $s$-dependence in the high-energy fixed-angle regime. However, in order to compare with experimental data at finite energies, the asymptotics are not sufficient and one must specify the background metric also in the IR region. For this purpose we choose the simplest confining AAdS background, which is AdS space with a finite IR cutoff, namely the hard-wall model also used in reference \cite{Polchinski:2001tt}. Note that as in the Sakai--Sugimoto model \cite{Sakai:2004cn} mesons are incorporated without the need to modify the background, because in the quenched (large-$N$) approximation mesons do not back react. 

Despite the limitations of our approach, we will show that, when combined with input from phenomenology, its predictions qualitatively agree with the data from reference \cite{Bromberg:1983he} in the high-energy fixed-angle regime.\,\,In contrast, agreement in the Regge regime is lacking, which could have been anticipated, as will be discussed in \hyperref[sec:2_2_Ansatz]{Section 2.2}.

This paper is organised as follows.\,\,In \hyperref[sec:2_background]{Section 2}, we review our holographic model and ansatz for pion scattering amplitudes from reference \cite{Armoni:2024gqv}.\,\,In \hyperref[sec:3_experimental]{Section 3}, we present two models for indirectly extracting $\pi^{+} \pi^{-} \to \pi^{+} \pi^{-}$ scattering data from $\pi^{-} p \to \pi^{+} \pi^{-} n$ measurements. In \hyperref[sec:4_comparison]{Section 4}, we compare our predictions with data from reference \cite{Bromberg:1983he} and offer predictions for all other $2$-to-$2$ pion scattering processes.\,\,Finally, in \hyperref[sec:5_conclusions]{Section 5}, we outline our findings and directions for future research.\,\,Two appendices are also included.\,\,\hyperref[app:A_graphs]{Appendix A} provides further comparisons to data for the discussion in \hyperref[sec:4_1]{Section 4.1}, while \hyperref[app:B_validation]{Appendix B} presents evidence suggesting that the $\pi^{-} p \to \pi^{+} \pi^{-} n$ data in reference \cite{Bromberg:1983he} include a significant contribution from an underlying 4-point scattering process.
\subsection*{Conventions and Notation}

We use natural units ($c = \hbar = 1$) and the mostly-plus sign convention for the metric tensor. The metric tensor $g_{MN}$ is reserved for curved spaces, while $\eta_{\mu\nu}$ is the Minkowski metric. Uppercase Latin and lowercase Greek letters label Lorentz indices ranging from 0 to 4 and 0 to 3, respectively.\,\,We adopt the convention where all external momenta are incoming. For $2$-to-$2$ scattering with initial momenta $k^{(1)}$ and $k^{(2)}$ and final momenta $k^{(3)}$ and $k^{(4)}$, we define the Mandelstam variables as follows:
\begin{equation}
    s = -\Big(k^{(1)} + k^{(2)}\Big)^2, \qquad t = -\Big(k^{(2)} + k^{(3)}\Big)^2, \qquad u = -\Big(k^{(1)} + k^{(3)}\Big)^2.
\end{equation}
For massless pion scattering, $t$ and $u$ can be given in terms of $s$ and the scattering angle $\theta$:
\begin{equation} \label{eq:1_variables}
    t = -\frac{s}{2} \, (1 - \cos \theta), \qquad u = -\frac{s}{2} \, (1 + \cos \theta).
\end{equation}
\section{High-Energy Pion Scattering in Holography} \label{sec:2_background}

This section begins with a review of our holographic model from reference \cite{Armoni:2024gqv} in \hyperref[sec:2_1_model]{Section~2.1}. In \hyperref[sec:2_2_Ansatz]{Section 2.2}, we describe how the Polchinski--Strassler proposal \cite{Polchinski:2001tt} can be used to write an ansatz for pion scattering amplitudes consistent with the constituent counting rule \eqref{eq:1_CCR}, following references \cite{Bianchi:2021sug, Armoni:2024gqv}.\,\,We also discuss the behaviour of this ansatz in the Regge regime. In \hyperref[sec:2_3_form]{Section 2.3}, we precisely formulate our 4-point pion scattering amplitudes, addressing their regularisation and numerical evaluation.
\subsection{Holographic QCD Model} \label{sec:2_1_model}

Motivated by references \cite{Sakai:2004cn, Kuperstein:2008cq, PhysRevD.69.065020}, reference \cite{Armoni:2024gqv} considers a holographic QCD model in which mesons are described at low energies by a U$(N_f)$ gauge theory in a five-dimensional AAdS space, governed by the action:
\begin{align} \label{eq:2_action1}
    S &\sim -\frac{1}{2} \int d^{4}x dw \sqrt{-g} \, \text{tr} \Big[g^{MN} g^{PQ} F_{MP} F_{NQ}\Big] \nonumber \\
    &\sim -\int d^{4}x dw \sqrt{-g} \, \text{tr} \bigg[\frac{h(w)}{2} \, \eta^{\mu\nu} \eta^{\rho\sigma} F_{\mu\rho} F_{\nu\sigma} + k(w) \, \eta^{\mu\nu} F_{\mu w} F_{\nu w}\bigg],
\end{align}
where $F_{MN}$ is the gauge field strength tensor of a U$(N_f)$ gauge field $A_M$, the coordinates $x^M = (x^{\mu}, w)$ parameterise the AAdS space, and the metric is assumed to take the form:
\begin{equation} \label{eq:2_metric1}
    ds^2 = g_{MN} dx^M dx^N = a(w) \, \eta_{\mu\nu} dx^{\mu} dx^{\nu} + b(w) \, dw^{2}, \qquad -\infty < w < +\infty.
\end{equation}
The functions $h(w)$ and $k(w)$ appearing in the action \eqref{eq:2_action1} are defined as follows:
\begin{equation}
    h(w) = \frac{1}{a^2(w)}, \qquad k(w) = \frac{1}{a(w) \, b(w)}.
\end{equation}
Moreover, the metric components $a(w)$ and $b(w)$ are assumed to be non-singular, positive, and symmetric under $w \to -w$, and the metric \eqref{eq:2_metric1} is taken to approach the AdS metric as $w \to \pm \infty$.\,\,In the UV-region ($w \to \pm \infty$), $a(w)$ and $b(w)$ can then be expressed as follows:
\begin{equation} \label{eq:2_UV}
   a(w) \sim \frac{w^2}{R_5^2}, \qquad b(w) \sim \frac{R_5^2}{w^2}.
\end{equation}
The setup described above is consistent with confinement and chiral symmetry breaking.\footnote{For a top-down example of a similar background in string theory, see reference \cite{Kuperstein:2008cq}.}

To dimensionally reduce the action \eqref{eq:2_action1}, the field $A_M$ is decomposed into complete sets of modes $\{\psi_{n}(w)\}_{n \geq 1}$ and $\{\phi_{n}(w)\}_{n \geq 0}$:
\begin{equation} \label{eq:2_gauge}
    A_{\mu}(x^{\mu}, w) = \sum_{n=1}^{\infty} B_{\mu}^{(n)}(x^{\mu}) \psi_{n}(w), \quad A_{w}(x^{\mu}, w) = \varphi^{(0)}(x^{\mu}) \phi_{0}(w) + \sum_{n=1}^{\infty} \varphi^{(n)}(x^{\mu}) \phi_{n}(w).
\end{equation}
As in reference \cite{Armoni:2024gqv}, $B_{\mu}^{(n)}$ correspond to a tower of massive meson fields alternating between vector and axial-vector states, with the lightest state $B_{\mu}^{(1)}$ identified as the $\rho$-meson field. The zero mode $\varphi^{(0)}$ is identified as the massless pion field. It has the correct quantum numbers and is massless because of its role as the Nambu--Goldstone boson of chiral symmetry breaking. The remaining fields $\varphi^{(n)}$ for $n \geq 1$ correspond to the would-be Nambu--Goldstone bosons absorbed by $B_{\mu}^{(n)}$.

To obtain a canonically normalised four-dimensional action for our mesons, we require the wavefunctions $\psi_{n}(w)$ and $\phi_{n}(w)$ to satisfy the following orthonormality conditions:
\begin{align} \label{eq:2_ortho}
    \int dw \sqrt{-g} \, h(w) \psi_{n}(w) \psi_{m}(w) = \delta_{nm}, \qquad \int dw \sqrt{-g} \, k(w) \phi_{n}(w) \phi_{m}(w) = \delta_{nm}.
\end{align}
We also choose $\psi_{n}(w)$ and $\phi_{n}(w)$ for $n \geq 1$ to satisfy the following equations:
\begin{equation} \label{eq:2_wave_eq}
    \frac{1}{\sqrt{-g} \, h(w)} \partial_w \Big(\sqrt{-g} \, k(w) \partial_w \psi_{n}(w)\Big) = -m_n^2 \psi_{n}(w), \qquad \partial_w \psi_{n}(w) = m_n \phi_{n}(w),
\end{equation}
where $m_n^2$ can be shown to correspond to the masses of the mesons $B_{\mu}^{(n)}$ for $n \geq 1$, while $\phi_{0}(w)$ is chosen to satisfy:
\begin{equation} \label{eq:2_pion1}
    \partial_w \Big(\sqrt{-g} \, k(w) \phi_{0}(w)\Big) = 0.
\end{equation}
Given the equation above, the pion wavefunction, which will be needed to write an ansatz for our pion scattering amplitudes in the next subsection, is given as follows:
\begin{equation} \label{eq:2_pion2}
    \psi_{\pi}(w) \coloneqq \phi_{0}(w) = \frac{1}{\sqrt{-g} \, k(w)} \sim \frac{R_5^3}{|w|^3},
\end{equation}
where an explicit normalisation factor has been omitted and the final expression includes only the leading term in the UV-region.

Note that in \hyperref[sec:3_experimental]{Section 4}, when comparing our predictions to experimental data, we will use only the hard-wall model, which is a bottom-up approach.\,\,This model is related to the model introduced here by taking, in equation \eqref{eq:2_metric1},
\begin{equation} a(w) = \frac{1}{b(w)} = \frac{w^2}{R_5^2}
\end{equation}
everywhere, but restricting the radial coordinate to 
\begin{equation}
	w_0 < w < \infty
\end{equation}
where $w_0 \sim \Lambda_{\text{QCD}}^{-1} > 0$ acts as an IR cut-off.\footnote{Note that, thanks to parity, we will not need to consider separately the regions of negative and positive $w$.}
\subsection{Scattering Amplitude Ansatz} \label{sec:2_2_Ansatz}

The action \eqref{eq:2_action1} should be interpreted as the kinetic term for massless open strings, which forms part of the effective low-energy theory describing our mesons.\,\,Meanwhile, computing high-energy pion scattering amplitudes in our model is expected to involve the exchange of the infinite tower of string states that UV-complete the action \eqref{eq:2_action1}.\,\,In a related setting, Polchinski and Strassler \cite{Polchinski:2001tt} introduced an ansatz for dilaton scattering amplitudes that accounts for the complete string tower and reproduces the constituent counting rule \eqref{eq:1_CCR} in a five-dimensional AdS space with an IR cut-off (for motivation behind their proposal, see also reference \cite{Brower:2006ea}, particularly Section 2; for an explicit example of their proposal applied to a simple toy model in field theory, see Section 3.3 of reference \cite{Armoni:2024gqv}).

Generalising the original proposal \cite{Polchinski:2001tt}, reference \cite{Armoni:2024gqv} proposed an ansatz in the context of the holographic model from \hyperref[sec:2_1_model]{Section 2.1} for pion and $\rho$-meson scattering amplitudes in the high-energy fixed-angle limit, with the ansatz for 4-point pion scattering given by:
\begin{equation} \label{eq:2_PS1}
    \mathcal{A} \Big(k_{\mu}^{(i)}\Big) = \int dw \sqrt{-g} \times \widetilde{\mathcal S} \Big(p_M^{(i)}, \zeta_M^{(i)}\Big) \times \prod_{i=1}^4 \psi^{(i)}(w),
\end{equation}
where $k_{\mu}^{(i)}$ for $i = 1, \ldots, 4$ are the four-dimensional momenta of the scattered pions, $\psi^{(i)}(w)$ are the pion wavefunctions, and $\widetilde{\mathcal S}$ denotes a 4-point superstring amplitude $\mathcal{S}$ for U$(N_f)$ gauge bosons with five-dimensional momenta $p_M^{(i)}$ and polarisations $\zeta_M^{(i)}$ in Minkowski space, modified by the conditions (\textbf{i})--(\textbf{iii}).

\begin{enumerate} [label=(\textbf{\roman*})]
    \item The Minkowski metric $\eta_{MN}$ is replaced by the AAdS metric $g_{MN}$ from equation \eqref{eq:2_metric1}:
    \begin{equation} \label{eq:2_con1}
    \eta_{MN} \to g_{MN}.
    \end{equation}
    \item The momenta $p_{\mu}^{(i)}$ are replaced by the momenta $k_{\mu}^{(i)}$, whereas the components $p_w^{(i)}$ are replaced by covariant derivatives $-i \nabla_w^{(i)}$ that act only on the wavefunctions $\psi^{(i)}(w)$ with the same index $(i)$:
    \begin{equation}
    p_{\mu}^{(i)} \to k_{\mu}^{(i)}, \qquad p_w^{(i)} \to -i \nabla_w^{(i)},
    \end{equation}
    with $\nabla_w^{(i)}$ and $\nabla_w^{(j)}$ commuting for $i \neq j$.
    
    In reference \cite{Armoni:2024gqv}, following reference \cite{Polchinski:2001tt}, it was shown that $p_w^{(i)}$ contribute as subleading corrections in $\alp / R_5^2$ to the 4-point pion scattering amplitude $\mathcal{A}$ in the high-energy fixed-angle limit.\,\,We will therefore neglect them in our analysis.\footnote{Note that, for $\rho$-meson scattering, completely neglecting the components $p_w^{(i)}$ leads to a vanishing result. See reference \cite{Armoni:2024gqv} for further details.}
    \item The pion polarisations $\zeta_M^{(i)}$ are fixed to:
    \begin{equation} \label{eq:2_con3}
    \zeta_M^{(i)} \to \big(0, 0, 0, 0, \zeta_w^{(i)}\big),
    \end{equation}
    while the wavefunctions $\psi^{(i)}(w)$ are replaced as follows:
    \begin{align}
    \zeta_w^{(i)} \psi^{(i)}(w) \rightarrow \psi_{\pi}^{(i)}(w).
\end{align}
\end{enumerate}

Following reference \cite{Armoni:2024gqv}, we now review how the pion scattering amplitude $\mathcal{A}$ given by equation \eqref{eq:2_PS1} reproduces the constituent counting rule, focusing on tree-level scattering. To show this, it suffices to consider only the UV-region of the AAdS metric \eqref{eq:2_metric1}, and while the explicit form of the superstring scattering amplitude $\mathcal{S}$ in Minkowski space is also not required, it will be useful for our later analysis to consider the following expression \cite{Green:1987sp}:
\begin{equation} \label{eq:2_S4}
    \mathcal{S} \Big(p_M^{(i)}, \zeta_M^{(i)}\Big) = -\frac{g_s^2}{2} \, \frac{\Gamma \big(-\alp s\big) \, \Gamma \big(-\alp t\big)}{\Gamma \big(1 - \alp s - \alp t\big)} \, \mathrm{K} \Big(p_M^{(i)}, \zeta_M^{(i)}\Big),
\end{equation}
where $g_s$ is the string coupling constant, $\mathrm{K}$ is the kinematical factor (see equation (7.4.42) in reference \cite{Green:1987sp}), and the Chan--Paton factors, temporarily omitted above, will be restored in \hyperref[sec:2_3_form]{Section 2.3}.

Applying the conditions (\textbf{i})--(\textbf{iii}) from equations \eqref{eq:2_con1}--\eqref{eq:2_con3} to $\mathcal{S}$ and neglecting $p_w^{(i)}$, we obtain:
\begin{equation} \label{eq:2_S4_tilde}
    \widetilde{\mathcal S} \Big(p_{\mu}^{(i)}, \zeta_M^{(i)}\Big) = -\frac{g_s^2}{2} \, \frac{\Gamma \big(-\alp \tilde{s}\big) \, \Gamma \big(-\alp \tilde{t}\big)}{\Gamma \big(1 - \alp \tilde{s} - \alp \tilde{t}\big)} \, \widetilde{\mathrm K} \Big(p_{\mu}^{(i)}, \zeta_M^{(i)}\Big),
\end{equation}
with $\widetilde{\mathrm K}$ given by:
\begin{align} \label{eq:2_K4_tilde}
    \widetilde{\mathrm K} = -\frac{1}{4} \Big[\tilde{s} \tilde{t} \, \zeta^{(1)} \cdot \zeta^{(3)} \, \zeta^{(2)} \cdot \zeta^{(4)} + \tilde{s} \tilde{u} \, \zeta^{(2)} \cdot \zeta^{(3)} \, \zeta^{(1)} \cdot \zeta^{(4)} + \tilde{t} \tilde{u} \, \zeta^{(1)} \cdot \zeta^{(2)} \, \zeta^{(3)} \cdot \zeta^{(4)}\Big].
\end{align}
In equations \eqref{eq:2_S4_tilde} and \eqref{eq:2_K4_tilde}, all dot products are evaluated using the AAdS metric \eqref{eq:2_metric1}, and scalar quantities built from these contractions are denoted with tildes to distinguish them from their Minkowski space counterparts.\,\,For example, $\tilde{s}$ is defined as follows:
\begin{equation} \label{eq:2_s_tilde}
    \tilde{s} = -g^{\mu\nu} \Big(k_{\mu}^{(1)} + k_{\mu}^{(2)}\Big) \Big(k_{\nu}^{(1)} + k_{\nu}^{(2)}\Big) = -\frac{\eta^{\mu\nu}}{a(w)} \Big(k_{\mu}^{(1)} + k_{\mu}^{(2)}\Big) \Big(k_{\nu}^{(1)} + k_{\nu}^{(2)}\Big) = \frac{s}{a(w)},
\end{equation}
with analogous definitions for $\tilde{t}$ and $\tilde{u}$.

For the following analysis, it is useful to recall that the high-energy fixed-angle limit of a 4-point scattering amplitude corresponds to taking $s$ and $|t|$ to infinity simultaneously, while keeping their ratio fixed.

Finally, the leading $s$-dependence of $\mathcal{A}$ in the high-energy fixed-angle limit is dictated only by the UV-region ($w \to \pm \infty$) of the AAdS metric \eqref{eq:2_metric1}, since all other contributions are suppressed by the exponential softness of the superstring scattering amplitude $\widetilde{\mathcal S}$ \cite{Polchinski:2001tt, Armoni:2024gqv}. In this limit, the $s$-scaling of $\mathcal{A}$ agrees with the constituent counting rule \eqref{eq:1_CCR}:
\begin{equation} \label{eq:2_scaling}
    \mathcal{A} \Big(k_{\mu}^{(i)}\Big) \sim \frac{1}{s^2}  \int \frac{dw_*}{w_*^7} \, \mathrm{B} \bigg(-\frac{1}{w_*^2}, -\frac{t}{s} \frac{1}{w_*^2}\bigg),
\end{equation}
where we have included only contributions from the UV-region of the metric \eqref{eq:2_metric1}, inserted the pion wavefunctions from equation \eqref{eq:2_pion2}, changed variables to $w_* = w / (R_5 \sqrt{\alp s})$, neglected an overall constant prefactor, and did not keep track of the dependence on the scattering angle $\theta$ coming from the kinematical factor.\,\,To restrict the integral above to contributions coming only from the UV-region of the AAdS metric \eqref{eq:2_metric1}, a finite cut-off in $w_*$ can be imposed. We will briefly elaborate on the meaning of this cut-off in our holographic model from \hyperref[sec:2_1_model]{Section 2.1}.

As discussed above and in reference \cite{Armoni:2024gqv}, the leading $s$-dependence of the pion scattering amplitude $\mathcal{A}$ in the high-energy fixed-angle limit is determined only by the UV-region of the metric \eqref{eq:2_metric1}.\,\,Consequently, the angular dependence associated with this leading-order behaviour should also be determined solely by the UV-region.\,\,To extract this dependence, we adopt a simple approach, as already mentioned above.\,\,We introduce an IR cut-off in the integral \eqref{eq:2_PS1} and use only the AdS metric.\,\,For this approach to be valid, the results should be independent of the IR cut-off, provided it is finite and non-zero.\,\,As we will show in \hyperref[sec:2_regular]{Section 2.3.1} based on numerical calculations, in the strict high-energy fixed-angle limit they are indeed independent of it.

On the other hand, in \hyperref[sec:4_1]{Section 4.1}, when comparing our predictions with experimental data, we will simply use the hard-wall model, where the cut-off is imposed from the beginning.

Now, to better inform our comparisons in \hyperref[sec:4_1]{Section 4.1}, we will examine the behaviour of the pion scattering amplitude $\mathcal{A}$ outside the high-energy fixed-angle regime in the context of the hard-wall model, while briefly commenting on why these results do not constitute genuine predictions of our holographic model from reference \cite{Armoni:2024gqv}.

In QCD, the well-known Regge regime of 4-point hadronic scattering amplitudes is characterised by $|t| \lesssim \Lambda^2 \ll s$, where $\Lambda$ is the QCD scale.\,\,In this regime, string theory scattering amplitudes in Minkowski space show Regge behaviour $s^{\alpha(t)}$ (where $\alpha(t)$ denotes a linear function of $t$), which is broadly consistent with experiments \cite{Green:1987sp}.\,\,There also exists a lesser-known regime, $\Lambda^2 \ll |t| \ll s$, in which string scattering amplitudes continue to exhibit Regge behaviour, whereas experimental data do not support such behaviour. In the following, we will restrict our attention to the broad regime $|t|/s \ll 1$, within which we will identify two distinct sub-regimes depending on the values of $t$ and $s$.\,\,Our analysis closely draws on the related discussion from reference \cite{Polchinski:2001tt, Brower:2006ea}.

Using standard Gamma function identities and neglecting the components $p_w^{(i)}$, we can express $\mathcal{A}$ as follows in the context of the model in the hard-wall AdS space, focusing on its $s$-scaling in the regime $|t| / s \ll 1$:
\begin{equation} \label{eq:2_Regge1}
    \mathcal{A} \Big(k_{\mu}^{(i)}\Big) \sim s \int_{w_0}^{\infty} \frac{dw}{w^7} \, \, \mathrm{B} \Big( -\alp \tilde{s}, -\alp \tilde{t} \Big)  \sim s \int_{0}^{\nu_0} d\nu \, \nu^{2} \bigg(\frac{|t|}{s}\bigg)^{\nu} \frac{\Gamma(\nu)}{\nu^{\nu}},
\end{equation}
where we retained only the leading $s$-dependence, did not keep track of overall factors of $t$, neglected the overall constant prefactor, and defined $\nu = \alp |t| R_5^2 / w^2$ and $\nu_0 = \alp |t| R_5^2 / w_0^2$.\,\,Notice that the integrand in the final expression differs from the integrand in the previous expression in the limit $w \to \infty$ (or equivalently $\nu \to 0^+$), though the resulting error from this is subleading in $s$.

To proceed, we group the terms in the integrand in equation \eqref{eq:2_Regge1} into two functions:
\begin{equation}
    f(\nu) = \nu^2 \, \bigg(\frac{|t|}{s}\bigg)^{\nu}, \qquad g(\nu) = \frac{\Gamma(\nu)}{\nu^\nu}.
\end{equation}
The function $f(\nu) \to 0$ as $\nu \to 0$ and $\nu \to \infty$, and attains an extremum at $\nu_* = 2/\ln(s/|t|)$, while the function $g(\nu) \to \infty$ as $\nu \to 0^+$ and decreases monotonically to zero as $\nu \to \infty$. Finally, the entire integrand in equation \eqref{eq:2_Regge1} tends to zero both as $\nu \to 0$ and $\nu \to \infty$, and possesses an extremum that is essentially set by the location of the extremum of $f(\nu)$.\footnote{Strictly speaking, notice that the limit $\nu \to \infty$ lies outside the integration domain in equation \eqref{eq:2_Regge1}. However, since we will later make use of the saddle point analysis, it is worthwhile to note that the integrand vanishes in this limit, allowing us to proceed with the standard analysis.} A more precise evaluation shows that the extremum shifts slightly, although it remains of the same order, with the leading approximation giving $\nu_* \sim 1 / \ln(s / |t|)$.\,\,Moreover, it can be shown that this extremum is a local maximum.

We are now in a position to recover the leading $s$-scaling of $\mathcal{A}$ in the regime $|t| / s \ll 1$. Before proceeding, it is worth noticing that both the local maximum $\nu_*$ and the upper cut-off $\nu_0$ of the integral in equation \eqref{eq:2_Regge1} depend on $t$ and $s$.\,\,Hence, whenever $\nu_0 < \nu_*$, the local maximum will lie outside of the integration domain.

Now, when the local maximum $\nu_*$ lies outside the integration domain, one may expect the integral in equation \eqref{eq:2_Regge1} to be dominated by the boundary value at $\nu_0$.\,\,This intuition can be made clear by expanding the integrand about $\nu = \nu_0$ and performing the resulting integral as follows:
\begin{align} \label{eq:2_Regge2}
    \mathcal{A} \Big(k_{\mu}^{(i)}\Big) \sim s \int_{0}^{\nu_0} d\nu \, \nu^{2} \bigg(\frac{|t|}{s}\bigg)^{\nu} \frac{\Gamma(\nu)}{\nu^{\nu}} \sim \frac{s^{1-\nu_0}}{\ln{|t|/s}},
\end{align}
where we omitted all $s$-independent factors in the leading-order term and neglected subleading terms in $1/\ln(s/|t|)$.\,\,In agreement with reference \cite{Polchinski:2001tt}, we recover Regge behaviour in the regime where $\nu_0 < \nu_*$, up to a logarithmic factor.\,\,In our case, $\mathcal{A} \sim s^{\alpha(t)}$ with $\alpha(t) = 1 + \alp t R_5^2 / w_0^2$, where $\alp R_5^2 / w_0^2$ can be identified as the effective Regge slope. Note that the intercept we obtain differs from the phenomenologically accepted value, hence we do not anticipate a good agreement with the data at small angles. 

Next, whenever the local maximum $\nu_{*}$ lies within the integration domain, the leading $s$-scaling of $\mathcal{A}$ can be found by using standard saddle point analysis:
\begin{equation} \label{eq:2_Regge3}
    \mathcal{A} \Big(k_{\mu}^{(i)}\Big) \sim s \int_{0}^{\infty} d\nu \, \nu^{2} \bigg(\frac{|t|}{s}\bigg)^{\nu} \frac{\Gamma(\nu)}{\nu^{\nu}} \sim \frac{s}{(\ln |t| / s)^2} \bigg(\frac{|t|}{s}\bigg)^{-\ln{\frac{|t|}{s}}} \bigg(-\ln{\frac{|t|}{s}}\bigg)^{-1/\ln{\frac{|t|}{s}}},
\end{equation}
where we also omitted all $s$-independent factors in the leading-order term, and extended $\nu_0 \to \infty$, noting that the resulting error in doing so is subleading in $s$.

The results in equations \eqref{eq:2_Regge2} and \eqref{eq:2_Regge3} are consistent with the corresponding results of reference \cite{Polchinski:2001tt} for scalar glueball scattering amplitudes, up to logarithmic factors. Now, in the context of the hard-wall model, the pion scattering amplitude $\mathcal{A}$ exhibits Regge behaviour at fixed, large $s$ and lower $|t|$ (as set by $\nu_0 < \nu_*$), while this behaviour is lost at higher $|t|$. As noted in reference \cite{Polchinski:2001tt} and references therein, the transition between these regimes is likely affected by multiloop effects. Given this, our tree-level approach is unlikely to reliably describe the Regge regime in a broad sense ($|t|/s \ll 1$) without distinguishing between the two regimes.

It is also worth pointing out that, following reference \cite{Brower:2006ea}, we do not expect that neglecting the components $p_w^{(i)}$ throughout the entire broad regime $|t| / s \ll 1$ is valid.

Now, let us also explain why these results likely do not follow through as genuine predictions of our holographic model from reference \cite{Armoni:2024gqv}. The issue is that the analysis above relies solely on the AdS metric, and we have seen that our results depend on features such as the IR cut-off in the hard-wall AdS model. In this sense, it would be inconsistent to consider only the UV region of metric \eqref{eq:2_metric1}; if the leading behaviour depends on the IR cut-off, the behaviour of the metric away from the UV region must also be taken into account.
\subsection{Precise Formulation of Scattering Amplitudes} \label{sec:2_3_form}

In this subsection, we present our model for 4-point pion scattering in a form suitable for comparison with experiment.\,\,We begin by restating the amplitude $\mathcal{A}$ from equation \eqref{eq:2_PS1}, paying particular attention to the dimensionful parameters.\,\,As mentioned in \hyperref[sec:2_2_Ansatz]{Section 2.2}, the leading $s$-dependence in the high-energy fixed-angle limit is governed by the UV-region of the AAdS metric \eqref{eq:2_metric1}.\,\,To isolate this contribution, we evaluate all expressions using the AdS metric and insert an IR cut-off $w_0$ on the integral in equation \eqref{eq:2_PS1}.\,\,As we will show, all dimensionful parameters, including the AdS radius $R_5$, the IR cut-off $w_0$, and the bare Regge slope $\alpha^{\prime}$ (except for the overall prefactor), can be combined into a single parameter, which we call the effective Regge slope $\tilde{\alpha}^{\prime}$, defined below, and which is distinct from the bare slope $\alpha^{\prime}$.

The main building block of our pion scattering amplitudes is given by:
\begin{equation} \label{eq:2_Amain}
    \mathcal{A}(s, t) \propto  \frac{s^2 + st + t^2}{s + t} \int_{w_0}^{\infty} dw \, w^{-7} \, \mathrm{B} \bigg(-\frac{\alp R_5^2}{w^2} \, s, -\frac{\alp R_5^2}{w^2} \, t\bigg),
\end{equation}
where an overall constant prefactor has been omitted.\,\,To proceed, we now introduce the following dimensionless integration variable:
\begin{equation}
    y \coloneqq \frac{\alp R_5^2}{w^2} \, s,
\end{equation}
with which equation \eqref{eq:2_Amain} can be written as follows:
\begin{equation} \label{eq:2_A1_y}
    \mathcal{A}(s, t) = \mathcal{C} \times \frac{s^2 + st + t^2}{s^3 u} \int_{0}^{y_0} dy \, y^2 \, \mathrm{B} \bigg(-y, -\frac{t}{s} \, y\bigg).
\end{equation}
Thus, the amplitude depends on dimensionful parameters through just two combinations. The first is the overall prefactor $\mathcal{C}$, as yet undetermined, and the second is the upper limit of the integral \eqref{eq:2_A1_y}, given by:
\begin{equation}
    y_0 = \frac{\alp R_5^2}{w_0^2} \, s \eqqcolon \tilde{\alpha}^{\prime} s.
\end{equation}
In the strict high-energy fixed-angle limit $y_0\to\infty$, our predictions do not depend on the exact value of $\tilde{\alpha}^{\prime}$, as long as it remains finite. Hence, the leading behavior does not depend on the value of the IR cut-off, or more generally, on how confinement is realized in the model.

When examining pion scattering, we must also account for the pion's isospin (realized in our holographic model using Chan--Paton factors, which we omitted earlier in \hyperref[sec:2_2_Ansatz]{Section 2.2}). To do this, we follow the standard procedure \cite{Ecker:1994gg, Igi:1998gn}, in which amplitudes of various pion scattering processes are represented by a single analytic function, the \emph{invariant amplitude}:
\begin{equation} \label{eq:2_Inv}
    \mathcal{I}(s, t) \coloneqq \frac{1}{2} \Big[\mathcal{A}(s, t) + \mathcal{A}(s, u) - \mathcal{A}(t, u)\Big].
\end{equation}
In terms of this function, the amplitude for any process $\pi^{i} \pi^{j} \to \pi^{k} \pi^{l}$ can be written as:
\begin{equation} \label{eq:2_Mijkl}
    \mathcal{M}_{ijkl}(s, t) = \mathcal{I}(s, t) \, \delta_{ij} \delta_{kl} + \mathcal{I}(t, s) \, \delta_{ik} \delta_{jl} + \mathcal{I}(u, t) \, \delta_{il} \delta_{jk},
\end{equation}
where $i \sim l = 1, 2, 3$ are the isospin indices, with $\pi^{\pm} = \frac{1}{\sqrt{2}} (\pi^1 \pm i \pi^2)$ and $\pi^0 = \pi^3$.

The invariant amplitude \eqref{eq:2_Inv} is composed of three terms.\,\,However, the term $\mathcal{A}(s, u)$ is essentially the same as $\mathcal{A}(s, t)$, as they are related by a reflection of the scattering angle, $\theta \to \pi - \theta$, or equivalently $\cos \theta \to -\cos \theta$.\,\,The term $\mathcal{A}(t, u)$ is given by:
\begin{equation} \label{eq:2_A2_y}
   \mathcal{A}(t, u) = \mathcal{C} \times \frac{s^2 + st + t^2}{s^4} \int_{0}^{y_0} dy \, y^2 \, \mathrm{B} \bigg(-\frac{u}{s} \, y, -\frac{t}{s} \, y\bigg),
\end{equation}
with the same constant $\mathcal{C}$ as in equation \eqref{eq:2_A1_y}.

To determine the angular dependence of all of our pion scattering amplitudes, we only need to compute two independent functions defined by the following expressions:
\begin{align}
    I^{(1)}(s, \theta) &= \int_{0}^{\tilde{\alpha}^{\prime} s} dy \, y^2 \, \mathrm{B} \bigg(-y, \frac{1}{2}(1 - \cos \theta) y\bigg), \label{eq:2_I1} \\ 
    I^{(2)}(s, \theta) &= \int_{0}^{\tilde{\alpha}^{\prime} s} dy \, y^2 \, \mathrm{B} \bigg(\frac{1}{2}(1 - \cos \theta) y, \frac{1}{2}(1 + \cos \theta) y\bigg), \label{eq:2_I2}
\end{align}
where we have used the kinematics defined earlier in equation \eqref{eq:1_variables} to express $t$ and $u$ in terms of $s$ and $\theta$.

Finally, all the relevant invariant amplitudes are given as follows:
\begin{align}
    \mathcal{I}(s, t) &= \frac{st + tu + su}{2 s^3} \Bigg[\frac{I^{(1)}(s, \theta)}{u} + \frac{I^{(1)}(s, \pi - \theta)}{t} - \frac{I^{(2)}(s, \theta)}{s}\Bigg], \\
    \mathcal{I}(t, s) &= \frac{st + tu + su}{2 s^3} \Bigg[\frac{I^{(1)}(s, \theta)}{u} - \frac{I^{(1)}(s, \pi - \theta)}{t} + \frac{I^{(2)}(s, \theta)}{s}\Bigg], \\
    \mathcal{I}(u, t) &= \frac{st + tu + su}{2 s^3} \Bigg[\frac{I^{(2)}(s, \theta)}{s} + \frac{I^{(1)}(s, \pi - \theta)}{t} - \frac{I^{(1)}(s, \theta)}{u}\Bigg].
\end{align}
We can also write these expressions as functions of $s$ and $\theta$:
\begin{align}   
    \mathcal{I}(s, t) &= -\frac{3 + \cos^{2} \theta}{s^2} \Bigg[\frac{I^{(1)}(s, \theta)}{1 + \cos \theta} + \frac{I^{(1)}(s, \pi - \theta)}{1 - \cos \theta} + \frac{I^{(2)}(s, \theta)}{2}\Bigg], \\
    \mathcal{I}(t, s) &= -\frac{3 + \cos^{2} \theta}{s^2} \Bigg[\frac{I^{(1)}(s, \theta)}{1 + \cos \theta} - \frac{I^{(1)}(s, \pi - \theta)}{1 - \cos \theta} - \frac{I^{(2)}(s, \theta)}{2}\Bigg], \\
    \mathcal{I}(u, t) &= -\frac{3 + \cos^{2} \theta}{s^2} \Bigg[-\frac{I^{(1)}(s, \theta)}{1 + \cos \theta} + \frac{I^{(1)}(s, \pi - \theta)}{1 - \cos \theta} - \frac{I^{(2)}(s, \theta)}{2}\Bigg].
\end{align}
\subsubsection{Regularisation and Numerical Methods} \label{sec:2_regular}

When integrating the expression in equation \eqref{eq:2_I1} with respect to the radial coordinate~$y$, singularities appear at all points where $y = n$ for positive integers $n$.\,\,The total number of such singularities is determined by the upper limit of the integral $y_0 = \tilde{\alpha}^{\prime} s$ and is given by the number of integers satisfying $n < \tilde{\alpha}^{\prime} s$.\,\,All of these singularities can be regularised, except for the case when the upper limit itself is a positive integer, i.e. when $\tilde{\alpha}^{\prime} s = n$.\,\,However, in the strict high-energy fixed-angle limit, where our approach applies, we only find an infinite set of regularisable singularities, since $s$ and $|t| \to \infty$ with their ratio held fixed.

As mentioned, evaluating equation \eqref{eq:2_I1} at integer values of $\tilde{\alpha}^{\prime} s$ leads to singularities that cannot be regularised.\,\,These can be shown to correspond to logarithmic $s$-channel singularities \cite{Bianchi:2021sug}, although we do not expect that these singularities are physical, as they are restricted to finite $s$.\,\,On the other hand, as long as $s$ does not coincide with a singularity, i.e.\,\,when $\tilde{\alpha}^{\prime} s \neq n$, all other singularities can be regularised to give a well-defined integral, from which the angular dependence of the amplitude can be computed.\,\,One way to achieve this is by shifting all the poles by $i \epsilon$, effectively giving a small width to all the mesons on the Regge trajectory.\footnote{Phenomenologically, we expect finite widths, since all of the states on the $\rho$-meson Regge trajectory are unstable and decay.\,\,In fact, the string decay width is expected to increase with spin, being proportional to the string length, which scales linearly with the mass \cite{Dai:1989cp, Peeters:2005fq, Iengo:2006gm, Sonnenschein:2017ylo}.\,\,In the 't Hooft limit, the excited states are stable \cite{Witten:1979kh}, since the width is $\sim 1/N$, and the width functions serve only as regulators, which are eventually taken to zero.}\,\,A finite physical limit can then be defined by taking the limit $\epsilon \to 0$ in an appropriate manner.

To implement this prescription, we write:
\begin{equation} \label{eq:2_A_epsilon}
    \mathcal{A}_{\epsilon}(s,t) \coloneqq \mathcal{C} \times \frac{s^2 + s t + t^2}{s^3 u} \int_0^{y_0} dy \, y^2 \, \mathrm{B} \Big(-y(1 - i \epsilon), -\frac{t}{s} \, y(1 - i \epsilon)\Big),
\end{equation}
which is finite for any real, non-zero $\epsilon$ and possess a non-singular limit as $\epsilon \to 0^{-}$ or $0^{+}$, except for the singularities at $\tilde{\alpha}^{\prime} s = n$.\,\,The limits $\epsilon \to 0^{-}$ and $0^{+}$ of equation \eqref{eq:2_A_epsilon} carry non-zero imaginary contributions, whose sign depends on the sign of $\epsilon$.

Now, since our predictions are based on tree-level superstring scattering amplitudes, the pion scattering amplitude $\mathcal{A}$ must be real.\,\,This follows from the unitarity constraint \cite{Peskin:1995ev}:
\begin{equation}
    2 \, {\rm Im} \, T= T \hspace{0.075em} T^{\dagger},
\end{equation}
where $T$ is the transfer matrix related to $\mathcal{A}$ via the LSZ formula, so $\mathcal{A} \sim T$.\,\,At tree level, the right-hand side above vanishes, implying that unitarity forces $\mathcal{A}$ to be real. This connection between unitarity and the reality of the amplitude can also be seen from the fact that, at large-$N$, the poles of the amplitude must lie on the positive real axis \cite{Caron-Huot:2016icg}.

To enforce this reality condition, we define the following regularised amplitude:
\begin{equation}
    \mathcal{A}^{(\text{reg})}(s, t) = \frac{1}{2} \bigg[\lim_{\epsilon \to 0^{-}} \mathcal{A}_{\epsilon}(s, t) + \lim_{\epsilon \to 0^{+}} \mathcal{A}_{\epsilon}(s, t)\bigg],
\end{equation}
which is equivalent to taking only the real-valued part of the one-sided limit $\epsilon \to 0^{-}$ or $0^{+}$ of equation \eqref{eq:2_A_epsilon}.\,\,The procedure described above is equivalent to the Cauchy principal value method for regularising the integral.

To proceed, we define $I_{n}^{(1)}(\theta)$ as the principal value of our integral around each pole. Denoting the integrand of equation \eqref{eq:2_I1} as $i^{(1)}(y,\theta)$, we calculate, for integer $n \geq 1$:
\begin{equation}
    I_{n}^{(1)}(\theta) = \lim_{\varepsilon \to 0^{+}} \Bigg[\int_{n - \frac{1}{2}}^{n - \varepsilon} dy \, i^{(1)}(y, \theta) + \int_{n + \varepsilon}^{n + \frac{1}{2}} dy \, i^{(1)}(y, \theta)\Bigg],
\end{equation}
and the regular integral for $n = 0$:
\begin{equation}
    I_0^{(1)}(\theta) = \int_0^{\frac{1}{2}} dy \, i^{(1)}(y, \theta).
\end{equation}
To obtain the $s \to \infty$ limit, we need to evaluate:
\begin{equation}
    I^{(1)}(\theta) = \sum_{n = 0}^{\infty} I^{(1)}_n(\theta).
\end{equation}
For finite $s$, we should cut-off the sum at $n_s = \lfloor \tilde{\alpha}^{\prime} s \rfloor$ and adjust the result to account for the difference in the integral from $n_s + \frac{1}{2}$ to the precise upper limit of $y_0 = \tilde{\alpha}^{\prime} s$.

In practice, summing over a finite number of terms suffices to converge to the $s \to \infty$ limit with high accuracy, as the beta function in the integrand is exponentially suppressed at large $y$.\,\,The number of terms that we need to sum depends on the value of the angle $\theta$. At finite angles, the sum is seen to converge quite rapidly, while near the edges, at angles close to $0$ or $\pi$, convergence is slower.\,\,This phenomenon will govern the energy dependence of the amplitude at large but finite $s$.\,\,At angles near $\theta = \frac{\pi}{2}$, where both $|t|$ and $|u|$ are large for large but finite $s$, the amplitude converges quickly to its asymptotic form.

\begin{figure}[h!]
    \centering
    \begin{subfigure}{0.45\textwidth} 
        \centering
        \includegraphics[width=\textwidth]{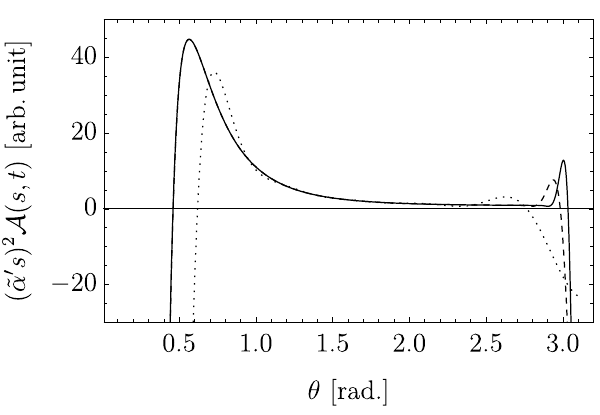}
        \caption{$\displaystyle (\tilde{\alpha}^{\prime} s)^2 {\cal A}(s, t)$}
    \end{subfigure} 
    \hfill
    \begin{subfigure}{0.45\textwidth}
        \centering
        \includegraphics[width=\textwidth]{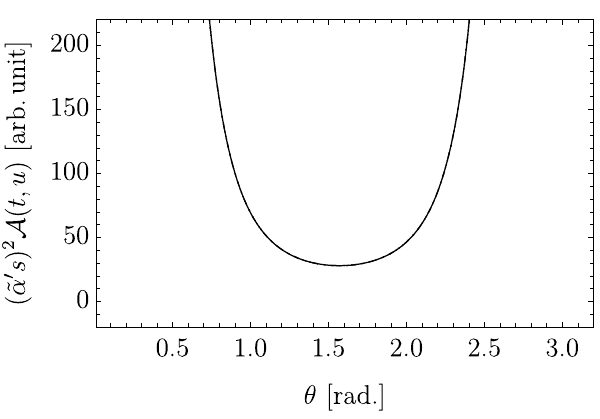}
        \caption{$\displaystyle (\tilde{\alpha}^{\prime} s)^2 {\cal A}(t, u)$}
    \end{subfigure}
    \caption{Angular dependence of the two independent amplitudes $(\tilde{\alpha}^{\prime} s)^2 {\cal A}(s, t)$ and $(\tilde{\alpha}^{\prime} s)^2 {\cal A}(t, u)$. The amplitudes are evaluated as functions of the scattering angle $\theta$ at various values of $\tilde{\alpha}^{\prime} s$ using equations \eqref{eq:2_A1_y} and \eqref{eq:2_A2_y} with the methods described above. The graphs show three lines: a dotted line for $\tilde{\alpha}^{\prime} s = 10.5$, a dashed line for $\tilde{\alpha}^{\prime} s = 50.5$, and a solid line for $\tilde{\alpha}^{\prime} s = 100.5$. In panel (b), all three lines are superimposed.  \label{fig:pion_plots}}
\end{figure}

To illustrate this behaviour, \hyperref[fig:pion_plots]{Figure 1} shows the computed angular dependence of the amplitudes $\mathcal{A}(s, t)$ and $\mathcal{A}(t, u)$ for various values of $\tilde{\alpha}^{\prime} s$. More precisely, the plots show $(\tilde{\alpha}^{\prime} s)^2 \mathcal{A}$ in the $s \to \infty$ limit, highlighting how the angular dependence reaches an asymptotic form as $s \to \infty$. The curves computed at $\tilde{\alpha}^{\prime} s = 10.5, 50.5,$ and $100.5$ demonstrate that the amplitude approaches a universal angular form in the high-energy limit. Consequently, based on these numerical calculations we expect that in the strict high-energy fixed-angle regime, our predictions become independent of the precise value of $\tilde{\alpha}^{\prime}$, provided it remains finite and non-zero.

The convergence to this asymptotic form is evident in both panels. In panel (a), the curves for $\tilde{\alpha}^{\prime} s = 50.5$ and $100.5$ are nearly indistinguishable at our plotting resolution, with residual difference confined to the region $\theta \sim \pi$ a discrepancy that we expect vanishes as $s \to \infty$. In panel (b), all three curves are completely superimposed at our plotting resolution, indicating faster convergence for the $\mathcal{A}(t, u)$ amplitude.

\clearpage
\section{Background Material on Experimental Data} \label{sec:3_experimental}

In this section, we describe how the data for $\pi^{+} \pi^{-} \to \pi^{+} \pi^{-}$ scattering can be extracted, under certain approximations, from the process $\pi^{-} p \to \pi^{+} \pi^{-} n$ studied in reference \cite{Bromberg:1983he}. To this end, we introduce two models of increasing generality for extracting the data that, together with our holographic model, will form the basis for the analysis in \hyperref[sec:4_comparison]{Section 4}.

\begin{figure}[h!]
    \centering
    \begin{subfigure}[b]{0.45\textwidth}
        \centering
        \includegraphics[width=0.65\textwidth]{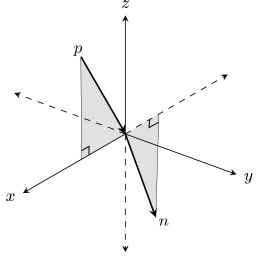}
        \caption{Proton and Neutron}
    \end{subfigure}
    \hfill
    \begin{subfigure}[b]{0.45\textwidth}
        \centering
        \includegraphics[width=0.65\textwidth]{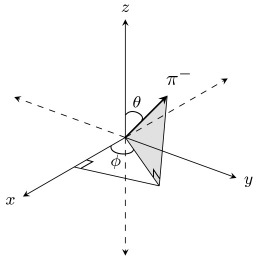}
        \caption{Outgoing Pion}
    \end{subfigure}
    \caption{Illustrations of the scattered particle momenta in the Gottfried--Jackson frame, obtained by boosting from the lab frame to the outgoing $\pi \pi$ rest frame.\,\,In this frame, the $z$-axis is along the incoming $\pi^{-}$ direction, $\hat{z} = \vec{q}^{\,\,(1)}/|\vec{q}^{\,\,(1)}|$, and the $y$-axis is given by $\hat{y} = \vec{q}^{\,\,(2)} \times \vec{q}^{\,\,(1)} / |\vec{q}^{\,\,(2)} \times \vec{q}^{\,\,(1)}|$, where the momenta $\vec{q}^{\,\,(1)}$ and $\vec{q}^{\,\,(2)}$ are defined in \hyperref[fig:q_def]{Figure 3}. The scattering angles $\theta$ and $\varphi$ are defined such that $-1 \leq \cos \theta \leq 1$ and $-\pi \leq \varphi \leq \pi$.}
    \label{fig:GJ_frame}
\end{figure}

In the experiment reported in reference \cite{Bromberg:1983he}, the process $\pi^{-} p \to \pi^{+} \pi^{-} n$ was studied at incident beam momenta $|\vec{q}_{\text{l.f.}}|$ of approximately 100\,GeV and 175\,GeV in the lab frame. The incident beam was primarily composed of pions (approximately 90\% at 100\,GeV and 92\% at 175\,GeV), while liquid hydrogen was used as the target.\,\,All the reported data were corrected for known experimental biases and acceptance losses, with uncertainties in these corrections remaining significantly smaller than the statistical errors across all data sets. Following the conventions of reference \cite{Bromberg:1983he}, we frequently use the Gottfried--Jackson frame in this section (see \hyperref[fig:GJ_frame]{Figure 2}).\,\,Additional information about the experiment can be found in references \cite{Fredericksen:1982hb, Stampke:1982cf, Bromberg:1983kv, Bromberg:1984ny}, while related studies are reported in references \cite{Aderholz:1964tpc, Aderholz:1964lgn, Bondar:1966zz, Hyams:1968zza, Robertson:1973tk, Grayer:1974cr, Birmingham-Rutherford-TelAviv-Westfield:1977ohi, Wicklund:1977in, Weilhammer:1979wg}.

\begin{figure}[h!]
    \centering
    \includegraphics[width=0.45\textwidth]{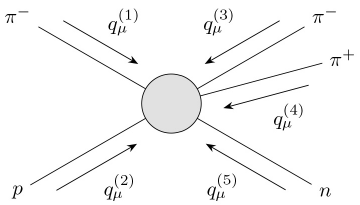}
    \caption{Schematic diagram of the process $\pi^{-} p \to \pi^{+} \pi^{-} n$ showing the four-momentum assignments for external particles, with all momenta taken as incoming by convention.}
    \label{fig:q_def}
\end{figure}

After averaging over the target proton spin and summing over the recoil neutron spin, the process $\pi^{-} p \to \pi^{+} \pi^{-} n$ depends on five independent kinematic variables.\,\,These can be chosen as $s$, $t_{pn}$, $|\vec{q}_{\text{l.f.}}|$, $\cos \theta$, and $\varphi$, as in reference \cite{Bromberg:1983he}.\,\,The scattering angles $\theta$ and $\varphi$ were introduced earlier in \hyperref[fig:GJ_frame]{Figure 2}, $|\vec{q}_{\text{l.f.}}|$ is the magnitude of the incoming beam momentum in the lab frame, while $s$ and $t_{pn}$ are defined as follows:
\begin{equation} \label{eq:3_var}
    s = -\Big(q_{\mu}^{(3)} + q_{\mu}^{(4)}\Big)^2, \qquad t_{pn} = -\Big(q_{\mu}^{(2)} + q_{\mu}^{(5)}\Big)^2,
\end{equation}
where the momenta $q_{\mu}^{(i)}$ for $i = 1, \ldots, 5$ were defined in \hyperref[fig:q_def]{Figure 3}.

\begin{figure}[h!]
    \centering
    \begin{subfigure}{0.45\textwidth}
        \centering
        \includegraphics[width=0.75\textwidth]{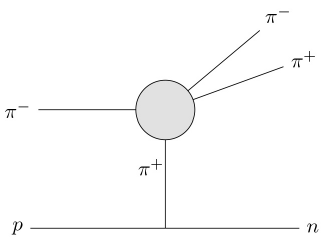}
        \caption{One-pion exchange (1PE)}
    \end{subfigure}
    \hspace{1cm}
    \begin{subfigure}{0.45\textwidth}
        \centering
        \includegraphics[width=0.75\textwidth]{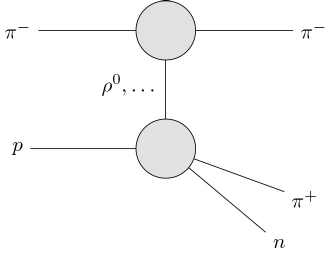}
        \caption{Isobar formation}
        \label{fig:isobar}
    \end{subfigure}
    \caption{Mechanisms contributing to the process $\pi^{-} p \to \pi^{+} \pi^{-} n$ at low $|t_{pn}|$.}
\end{figure}

Neglecting background contributions, to be discussed later, the process $\pi^{-} p \to \pi^{+} \pi^{-} n$ at low $|t_{pn}|$ has long been known to be approximated by effective one-pion exchange (1PE), as suggested early on in reference \cite{Chew:1958wd}.\,\,This approximation is motivated by the observation that the pion is the lightest strongly interacting particle.\,\,That said, a competing mechanism associated with isobar formation can also contribute to the same final state at low $|t_{pn}|$ (see \hyperref[fig:isobar]{Figure 4(b)}), obscuring an interpretation based on 1PE.\,\,However, experimental data at incoming beam momenta between 1 and 5\,GeV already indicate a 1PE-dominated region, with higher momenta expected to further suppress isobar contributions \cite{Martin:1976mb}.

Assuming 1PE dominance, the doubly inclusive differential cross section for the process $\pi^{-} p \to \pi^{+} \pi^{-} n$ is given by \cite{Martin:1976mb}:
\begin{equation} \label{eq:3_1PE}
    \frac{d^4 \sigma}{d\sqrt{s} \hspace{0.05em} dt_{pn} \hspace{0.05em} d\!\cos \theta \hspace{0.05em} d\varphi} = \frac{g^2}{32 \pi^3} \, \frac{(-t_{pn})}{(t_{pn} - m_{\pi}^2)^2} \, \frac{s \, |\vec{q}_{\pi\pi}|}{s_{\pi p} \, |\vec{q}_{\text{c.m.}}|^2} \, \frac{d\sigma_{\pi\pi}^{\text{{\it off}}}}{d\!\cos \theta},
\end{equation}
where $g^2 / 4\pi$ is the coupling for the phenomenological Yukawa interaction between the pion and nucleons at low energies, $m_{\pi}$ is the pion mass, $\vec{q}_{\pi\pi}$ is the final-state pion momentum in the $\pi\pi$ rest frame, $\vec{q}_{\text{c.m.}}$ is the incoming pion-proton system centre-of-mass momentum, $s_{\pi p}$ is defined as:
\begin{equation}
    s_{\pi p} = -\Big(q_{\mu}^{(1)} + q_{\mu}^{(2)}\Big)^2 = -\Big(q_{\mu}^{(3)} + q_{\mu}^{(4)} + q_{\mu}^{(5)}\Big)^2,
\end{equation}
and $\sigma_{\pi\pi}^{\text{{\it off}}}$ is the off-shell $\pi^{+} \pi^{-} \to \pi^{+} \pi^{-}$ cross section.\footnote{Details on how off-shell effects can be included can be found, for example, on page 33 in reference \cite{Martin:1976mb}.}

Although 1PE is expected to dominate the process $\pi^{-} p \to \pi^{+} \pi^{-} n$ at low $|t_{pn}|$, it can at best provide only a qualitative description.\,\,In particular, as observed in reference \cite{Stampke:1982cf}, the $\pi^{-} p \to \pi^{+} \pi^{-} n$ data from reference \cite{Bromberg:1983he} show deviations from equation \eqref{eq:3_1PE}, including a dependence on the angle $\varphi$, a non-vanishing behaviour as $|t_{pn}| \to 0$, and a steeper fall-off at larger $|t_{pn}|$ (see also reference \cite{Martin:1976mb}).\,\,As emphasised in reference \cite{Stampke:1982cf}, these discrepancies indicate the need to account for additional mechanisms beyond 1PE, such as heavier-meson exchanges, finite-size contributions, and absorption effects.

Among possible modifications to the 1PE model, called backgrounds in the literature, absorption effects are known to capture the most features of the $\pi^{-} p \to \pi^{+} \pi^{-} n$ data \cite{Martin:1976mb}. For this reason, we will focus on a modification of the 1PE model designed to incorporate the effects of pion absorption in the incoming and outgoing channels.\,\,While certainly not the only mechanism required for a complete description of the data, we expect these effects to provide a useful approximation beyond the simple 1PE model.

Specifically, we consider the Poor Man's Absorption (PMA) model, designed to provide a simplified description of the absorption effects mentioned above, with the doubly inclusive differential cross section for the process $\pi^{-} p \to \pi^{+} \pi^{-} n$ given by \cite{Martin:1976mb}:
\begin{equation} \label{eq:3_PMA1}
    \frac{d^{4} \sigma}{d\sqrt{s} \hspace{0.05em} dt_{pn} \hspace{0.05em} d\!\cos \theta \hspace{0.05em} d\varphi} = \frac{g^2}{4 \pi} \, \frac{s \, |\vec{q}_{\pi\pi}|}{m_p^2 \, |\vec{q}_{\text{l.f.}}|^2} \, \Big(P_0 + P_1 \cos \varphi\Big),
\end{equation}
with:
\begin{align} \label{eq:PMA2}
    P_0 &= \frac{-t_{pn}}{(t_{pn} - m_{\pi}^2)^2} \, F_0^2 \, \big|\mathcal{M}^{\text{{\it off}}}\big|^2 + \frac{s}{(s - m_{\pi}^2)^2} \, \big|C_A\big|^2 F_1^2 \, \bigg|\frac{\partial \mathcal{M}^{\text{{\it off}}}}{\partial \theta}\bigg|^2, \\
    P_1 &= \frac{(-t_{pn})^{1/2}}{(t_{pn} - m_{\pi}^2)} \frac{s^{1/2}}{(s - m_{\pi}^2)} \, \operatorname{Re}(C_A) F_0 F_1 \, \frac{\partial}{\partial \theta} \big|\mathcal{M}^{\text{{\it off}}}\big|^2,
\end{align}
where $m_p$ is the proton mass, $F_0$ and $F_1$ are collimating form factors that depend on $t_{pn}$, $C_A$ is a complex-valued, $s$-dependent absorption parameter, and $\mathcal{M}^{\text{{\it off}}}$ denotes the off-shell scattering amplitude for the process $\pi^{+} \pi^{-} \to \pi^{+} \pi^{-}$.

We now turn to the experimental data found in reference \cite{Bromberg:1983he}, where the observed $\pi\pi$ decay distributions from the process $\pi^{-} p \to \pi^{+} \pi^{-} n$ were reported in terms of a spherical harmonic expansion.\,\,For $|t_{pn}| < 0.15$\,GeV$^2$, only the $m = 0$ and $1$ moments were significant, allowing the $\pi\pi$ decay distribution at fixed $s$, $t_{pn}$, and $|\vec{q}_{\text{l.f.}}|$ to be approximated by:
\begin{equation} \label{eq:3_IP}
    I_{\text{P}}(\theta, \varphi) = I_{0}(\theta) + I_{1}(\theta) \cos \varphi,
\end{equation}
where $I_{\text{P}}$ is proportional to the doubly inclusive differential cross section for the process $\pi^{-} p \to \pi^{+} \pi^{-} n$, and $I_0$ and $I_1$ denote the $m = 0$ and $1$ moments, respectively.

In reference \cite{Bromberg:1983he}, data for $I_0$ were presented, while $I_1$ data were not always available. The $I_0$ data were obtained from fits performed on $s^{1/2}$ and $\cos \theta$ bins over the interval $|t_{pn}| < 0.15$\,GeV$^2$ (rather than at a single $|t_{pn}|$ value).\,\,Measurements were taken at both incoming beam momenta of $|\vec{q}_{\text{l.f.}}| \approx 100$\,GeV and $175$\,GeV, enabling a simple qualitative check of whether the studied process $\pi^{-} p \to \pi^{+} \pi^{-} n$ is dominated by a $|\vec{q}_{\text{l.f.}}|$-independent $\pi\pi$ production mechanism, successfully carried out in \hyperref[app:B_validation]{Appendix B}.\,\,Finally, by comparing the reported $I_0$ data with the 1PE or PMA model, we can obtain indirect information on the scattering process $\pi^{+} \pi^{-} \to \pi^{+} \pi^{-}$, as will be discussed in the next subsection.
\section{Comparisons and Predictions} \label{sec:4_comparison}

This section presents our model's predictions using the bottom-up hard-wall AdS model for high-energy pion scattering. In \hyperref[sec:4_1]{Section 4.1}, we compare our predictions for the angular dependence of $\pi^{+} \pi^{-} \to \pi^{+} \pi^{-}$ scattering with experimental data extracted from $\pi^{-} p \to \pi^{+} \pi^{-} n$ measurements reported in reference \cite{Bromberg:1983he}. In \hyperref[sec:4_2]{Section 4.2}, we present predictions for all $2$-to-$2$ pion scattering processes in one place.
\subsection{Comparisons to Data} \label{sec:4_1}

As outlined in \hyperref[sec:3_experimental]{Section 3}, the available data come from $\pi^{-} p \to \pi^{+} \pi^{-} n$ measurements and can be related to $\pi^{+} \pi^{-} \to \pi^{+} \pi^{-}$ scattering only with additional models.\,\,In our analysis, we use both of the models introduced earlier.\,\,From the discussion around equation \eqref{eq:3_IP}, we see that the available data consist of the components $I_0(\theta)$ and $I_1(\theta)$, which together form the $\pi\pi$ decay distribution equal to the doubly inclusive differential cross section for the process $\pi^{-} p \to \pi^{+} \pi^{-} n$ at fixed $s$, $t_{pn}$, and $|\vec{q}_{\text{l.f.}}|$ up to an overall normalisation constant \cite{Martin:1976mb}.

For comparisons using the 1PE model, only the $I_0(\theta)$ component can be considered, as the model does not account for $\varphi$ dependence, highlighting its inability to completely describe the process $\pi^{-} p \to \pi^{+} \pi^{-} n$.\,\,In contrast, the PMA model provides predictions for both $I_0(\theta)$ and $I_1(\theta)$.\,\,However, since the available data in reference \cite{Bromberg:1983he} include only $I_0(\theta)$, our comparison with the PMA model will also be limited to this component.

The fitting model for comparisons using the 1PE model at fixed $t_{pn}$ and $|\vec{q}_{\text{l.f.}}|$ will be:
\begin{equation}
    I_0(s, \theta) =\mathcal{C}_0\, \big|\mathcal{M}(s, \theta)\big|^2\,.
\end{equation}
The fitting model for comparisons using the PMA model at fixed $t_{pn}$ and $|\vec{q}_{\text{l.f.}}|$ will be:
\begin{equation}
    I_0(s, \theta) = \mathcal{C}_1 \, \Bigg[\big|\mathcal{M}(s, \theta)\big|^2 + \frac{\mathcal{C}_2}{ \tilde{\alpha}^{\prime} s} \bigg\vert\frac{\pa \mathcal{M}(s, \theta)}{\pa\theta}\bigg\vert^2\Bigg]\,.
\end{equation}
The fitting parameters ${\mathcal C}_0$ and ${\mathcal C}_1$ are expected to have an energy dependence, as can be read from equations \eqref{eq:3_1PE} and \eqref{eq:3_PMA1}--\eqref{eq:PMA2}. The parameter ${\mathcal C}_2$ is dimensionless and approximately independent of $s$.

The relevant pion scattering amplitude appearing in the equations above is given by:
\begin{equation}
    \mathcal{M}[\pi^{+} \pi^{-} \to \pi^{+} \pi^{-}] = \mathcal{I}(s,t) + \mathcal{I}(t,s) = \mathcal{A}(s,t),
\end{equation}
where $\mathcal{A}(s,t)$ is given by equation \eqref{eq:2_A1_y} after regularisation.

Apart from the fitting parameters $\mathcal{C}_0$ in the 1PE model and $\mathcal{C}_1$ and $\mathcal{C}_2$ in the PMA model, there is one meaningful dimensional parameter, the Regge slope $\tilde{\alpha}^\prime$. Instead of fitting it to the data, we may recall that $\tilde{\alpha}^\prime$  is the effective Regge slope in our approach, and thus we can set it to its phenomenological value:
\begin{equation}
    \tilde{\alpha}^{\prime} = 0.9 \, \text{GeV}^{-2},
\end{equation}
which is the slope of the $\rho$-meson linear Regge trajectory \cite{Collins:1977jy, Ebert:2009ub, Sonnenschein:2018fph}. 

\begin{figure}[ht!]
    \centering
    \begin{subfigure}[b]{0.45\textwidth}
         \includegraphics[width=\textwidth]{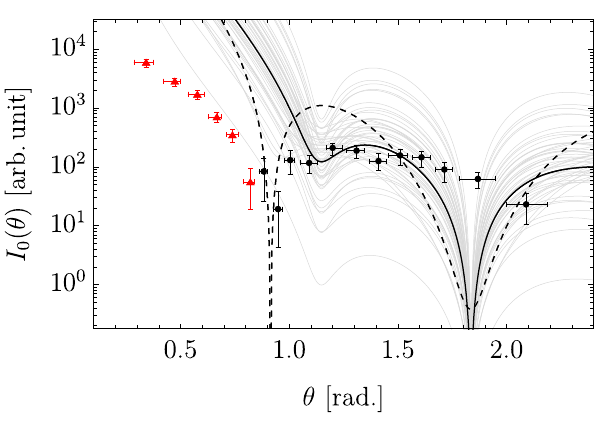}\caption{$\sqrt s = 2.1$--$2.3$ GeV} 
    \end{subfigure}
    \begin{subfigure}[b]{0.45\textwidth}
         \includegraphics[width=\textwidth]{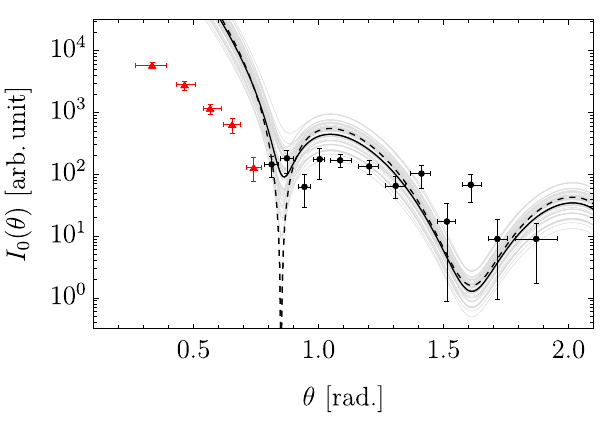}\caption{$\sqrt s = 2.3$--$2.5$ GeV}
    \end{subfigure}
    
    \vspace{0.5cm}
    
    \begin{subfigure}[b]{0.45\textwidth}
         \includegraphics[width=\textwidth]{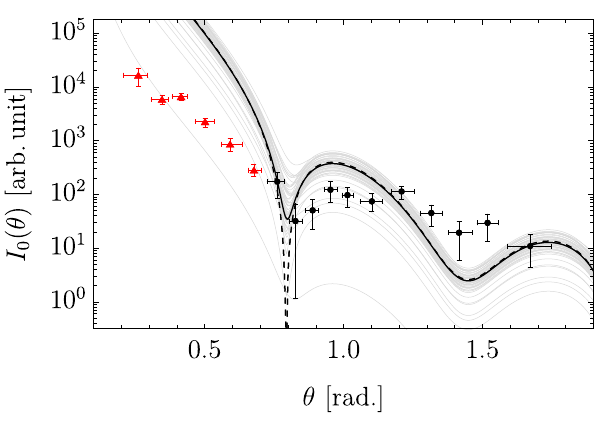}\caption{$\sqrt s = 2.5$--$2.7$ GeV} 
    \end{subfigure}
    \begin{subfigure}[b]{0.45\textwidth}
         \includegraphics[width=\textwidth]{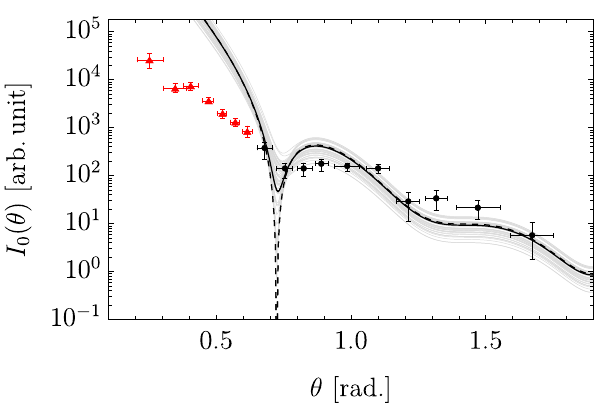}\caption{$\sqrt s = 2.7$--$3.0$ GeV}
    \end{subfigure}

    \vspace{0.5cm}
    
    \begin{subfigure}[b]{0.45\textwidth}
         \includegraphics[width=\textwidth]{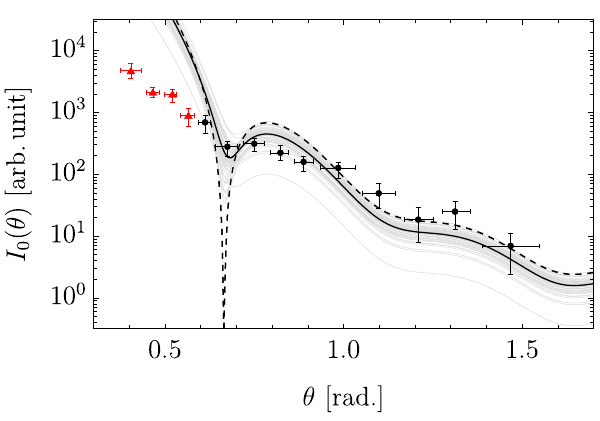}\caption{$\sqrt s = 3.0$--$3.3$ GeV} 
    \end{subfigure}
    \begin{subfigure}[b]{0.45\textwidth}
         \includegraphics[width=\textwidth]{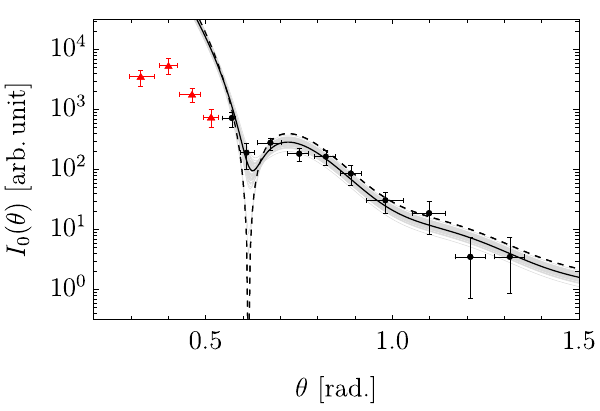}\caption{$\sqrt s = 3.3$--$3.6$ GeV}
    \end{subfigure}
    \caption{Fits for the 175 GeV dataset. The solid black line represents the best fit using the PMA model, while the grey band corresponds to one standard deviation of the fitting parameters in the PMA model. The dashed black line corresponds to a best fit using the 1PE model. Data points shown in red were excluded from the fit. The experimental data were digitised from Figure 29 in reference \cite{Bromberg:1983he}; it is worth noting that we were not able to recover all data points or errors faithfully, although such cases were rare. The $s$-values indicated above correspond to the binned $s$-values included in the experimental measurements, whereas our fits are made only to the average of these binned values.}
    \label{fig:fits175}
\end{figure}

In \hyperref[fig:fits175]{Figure 5}, we plot the fitting results for the 175 GeV data set. We can see that the best fits are, as expected, at the highest energies available, $\sqrt{s} = 3.15$ and $3.45$ GeV, where ${\tilde \alpha}^\prime s \approx 9$--10 is beginning to be large. These last two fit remarkably well at a qualitative level, which is encouraging despite the many approximations involved. The high-energy fits are better also in the sense that the errors in the fitting parameters are smaller, a sign that the fitting algorithm converges better with the given model. In \hyperref[tab:fitparams175]{Tables 1} and \hyperref[tab:fitparams100]{2}, we list the fitting parameters we obtained. Their precise values have no particular physical significance for our purposes, and we list them only for completeness. Additional plots, including those for the 100 GeV data set are in \hyperref[app:A_graphs]{Appendix A}.

A notable feature of the amplitude is a local minimum, or ``dip'', at $\tilde{\alpha}^{\prime} t \approx -1$, which appears at both lower and higher $s$-values.

The Minkowski space string amplitude predicts a large number of zeros and hence dips at $\tilde{\alpha}^{\prime} u = -n$ for any integer $n$. In the older literature, this feature, known as the Veneziano dip, was observed in various scattering experiments, though not always for all values of $n$ \cite{Odorico:1971dd, Odorico:1972vh, Pennington:1973dz, Eguchi:1973iv, Stampke:1982cf}.  It was argued to be a consequence of the Veneziano model, whose pole structure also implies an analogous structure of zeros responsible for the dips. Even though the basis for this phenomenon is the high-energy behaviour, it persists at low energies, as the analytic structure of the amplitude remains constrained.

On the other hand, the dip at $\tilde{\alpha}^{\prime} t \approx -1$ is a different feature from the Veneziano dips at constant $\tilde{\alpha}^{\prime} u$. Our holographic model, after the integration over the flat space amplitude that has many dips, has only one or two zeros, though remarkably the observed dip at $\tilde{\alpha}^{\prime} t \approx -1$ is retained. Its location also serves as an independent verification of the value we set for $\tilde{\alpha}^{\prime}$, as increasing or decreasing $\tilde{\alpha}^{\prime}$ will shift the dip to smaller or larger angles, respectively. The value ${\tilde \alpha}^\prime =0.9$ GeV${^{-2}}$ is consistent with the position of the dip in the data.

We expect that the dip eventually moves away from the point $\tilde{\alpha}^{\prime} t = -1$, as increasing $s$ pushes it to larger values of $|t|$. In the $s \to \infty$ limit plotted in \hyperref[fig:pion_plots]{Figure 1}, the zero is asymptotically at the finite angle $\theta \approx 0.46$, meaning that eventually the position of the dip moves to $|t| \to \infty$. One can check numerically where the zero is as a function of ${\tilde \alpha}^\prime  s$ and observe that at finite energies it stays approximately near ${\tilde \alpha}^\prime  t = -1$ for energies up to around ${\tilde \alpha}^\prime  s = 20$. Our model then predicts that at higher energies, there will be a dip in the scattering amplitude, but approaching a constant angle rather than a constant value of $t$.

An important consideration that we have to make when fitting the data is the mismatch in the Regge regime, causing a discrepancy with our model at low $t$, with the threshold being around $|t|/s \approx 0.1$. This will require us to adjust our fitting strategy accordingly. In light of this, our preferred fitting strategy was to fit all points to the right of the dip, and one or two points immediately to the left of it, even when those were below the small angle threshold of $|t|/s \approx 0.1$. 

Regardless of the choice of fitting strategy, we find general qualitative agreement between our model and the data, especially at the energies that start probing the high-energy fixed-angle regime where our model is expected to apply.

\begin{table}[h!]
    \centering
    \begin{tabular}{c||c|c|c}
     \multicolumn{4}{c}{} \\
        $\sqrt{s}$ [GeV] & $\mathcal{C}_0$ [arb.\,\,units] & $\mathcal{C}_1$ [arb.\,\,units] & $\mathcal{C}_2$ [dimensionless] \\
        \hline
        $2.1$--$2.3$ & $6930\pm4420$ & $770\pm1350$ & $1.251 \pm 2.730$ \\
        \hline
        2.3--2.5 & $3430\pm1530$ & $2760\pm1440$ & $0.007 \pm 0.014$\\
        \hline
        $2.5$--$2.7$ & $2420\pm940$ & $2290\pm1030$ & $0.003 \pm 0.011$\\
        \hline
         $2.7$--$3.0$ & $3970\pm910$ & $3770\pm1030$ & $0.003 \pm 0.008$\\ \hline
         $3.0$--$3.3$ & $6250\pm2370$ & $4120\pm1060$ & $0.010 \pm 0.007$\\ \hline
        $3.3$--$3.6$ & $5630\pm1780$ & $4060\pm530$ & $0.007 \pm 0.003$\\
    \end{tabular}
    \caption{Fitting parameters from comparisons to 175 GeV data sets.}
    \label{tab:fitparams175}
\end{table}

\begin{table}[h!]
    \centering
    \begin{tabular}{c||c|c|c}
     \multicolumn{4}{c}{} \\
        $\sqrt{s}$ [GeV] & $\mathcal{C}_0$ [arb.\,\,units] & $\mathcal{C}_1$ [arb.\,\,units ]& $\mathcal{C}_2$ [dimensionless] \\\hline
        $2.1$--$2.3$ & $7490\pm4060$ & $1180\pm1620$ & $0.840 \pm 1.509$\\
        \hline
        $2.3$--$2.5$ & $5290\pm2960$ & $4840\pm2980$ & $ \sim 0 \pm 0.009$\\ \hline
        $2.5$--$2.7$ & $3680\pm1500$ & $3480\pm1630$ & $\sim 0 \pm 0.013$\\\hline
         $2.7$--$3.1$ & $6720\pm1750$ & $6620\pm1990$ & $0.001 \pm 0.006$\\
    \end{tabular}
    \caption{Fitting parameters from comparisons to 100 GeV data sets.}
    \label{tab:fitparams100}
\end{table}

\clearpage

\subsection{Additional Predictions} \label{sec:4_2}

Our model provides concrete predictions for all possible channels of $2\to2$ pion scattering. We will present some of them here, discussing potential qualitative features which could be observed in future experiments.

To begin with, let us examine the amplitude for $\pi^+\pi^-\to\pi^+\pi^-$. In the previous section, we compared it with an indirect measurement coming from the experiment $\pi^{-} p \to \pi^{+} \pi^{-} n$, but here we plot the pion amplitude directly, computing  it over the whole range of the scattering angle and at various energies, higher than the data we examined.

\begin{figure} [h!]
    \centering
    \begin{subfigure}[b]{0.48\textwidth}
       \centering
       \includegraphics[width=\textwidth]{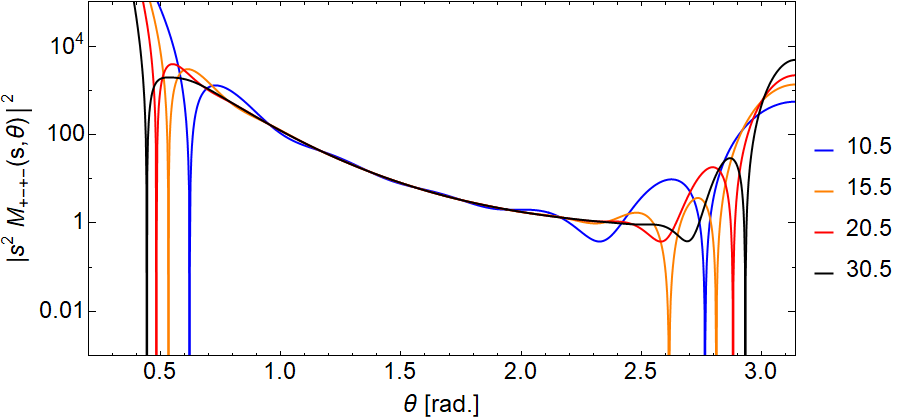}
    \end{subfigure}
    \hfill
    \begin{subfigure}[b]{0.48\textwidth}
        \centering
        \includegraphics[width=\textwidth]{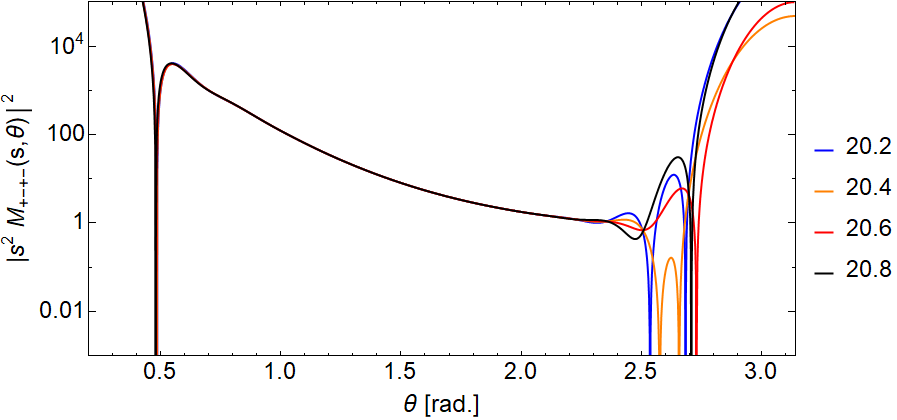}
    \end{subfigure}
    \caption{The angular dependence of the differential cross-section of the $\pi^+\pi^-\to\pi^+\pi^-$ amplitude. The legends to the right of the plots display the value of ${\tilde \alpha}^\prime  s$ for each curve.}
    \label{fig:predictions_pmpm}
\end{figure}

Aside from the dip at ${\tilde \alpha}^\prime t \approx -1$ discussed above, there are additional zeros of the amplitude at ${\tilde \alpha}^\prime  u \approx -1$ (which is consistent with the position of the Veneziano dip) and ${\tilde \alpha}^\prime  u \approx -0.4$. The latter, third zero is not present for all values of the energy. The positions of these additional minima are more dependent on the energy, as can be seen in figure \ref{fig:predictions_pmpm}. In particular, the right figure shows that the positions of these zeros are sensitive to the particular value of ${\tilde \alpha}^\prime  s$. They are outside the range of the experimental data of \cite{Bromberg:1983he}, and it would be an interesting feature to verify against other experiments. 

Note that in these and following figures we multiply the pion amplitude by $s^2$ so that we can compare directly the angular dependence of the different plots.

One can also see from the plots that the amplitude converges faster to the $s\to\infty$ limit when the angles are not small, i.e. when $|t|$ is also large, so we can already probe the high-energy limit at ${\tilde \alpha}^\prime s\sim 10$.

Moving on to the other processes, in total, there are five distinct processes of $2$-to-$2$ pion scattering. Other than the previously examined ${\cal M}_{+-+-}$, the remaining relevant amplitudes are ${\cal M}_{++++}$, ${\cal M}_{+0+0}$, ${\cal M}_{+-00}$, and ${\cal M}_{0000}$. Their explicit forms can be obtained from eq. \eqref{eq:2_Mijkl}. All other processes are directly related to one of these by symmetry.

\begin{figure}[hbt!]
    \centering
    \begin{subfigure}[b]{0.45\textwidth}
        \centering
        \includegraphics[width=\textwidth]{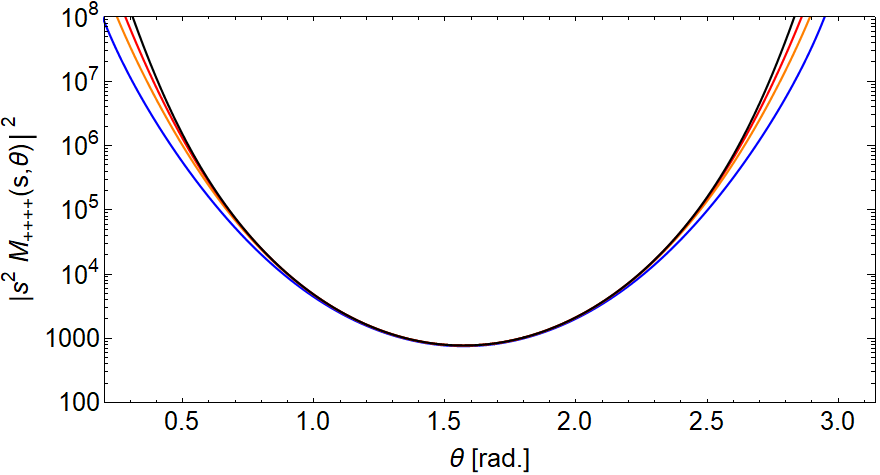}
        \caption{$|s^{2} {\cal M}_{++++}|^2$}
    \end{subfigure}
    \hfill
    \begin{subfigure}[b]{0.45\textwidth}
        \centering
        \includegraphics[width=\textwidth]{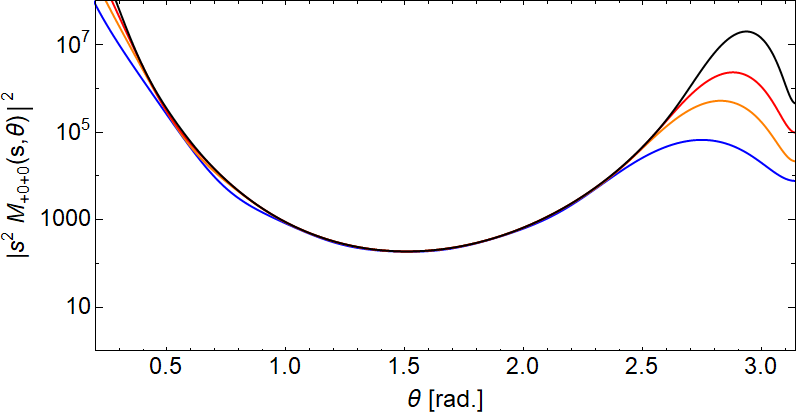}
        \caption{$|s^{2} {\cal M}_{+0+0}|^2$}
        \label{fig:7b}
    \end{subfigure}
    \vskip\baselineskip
    \begin{subfigure}[b]{0.45\textwidth}
        \centering
        \includegraphics[width=\textwidth]{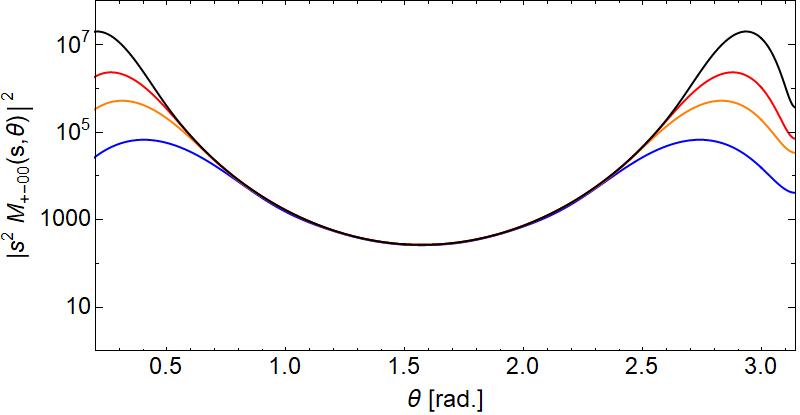}
        \caption{$|s^{2} {\cal M}_{+-00}|^2$}
        \label{fig:7c}
    \end{subfigure}
    \hfill
    \begin{subfigure}[b]{0.45\textwidth}
        \centering
        \includegraphics[width=\textwidth]{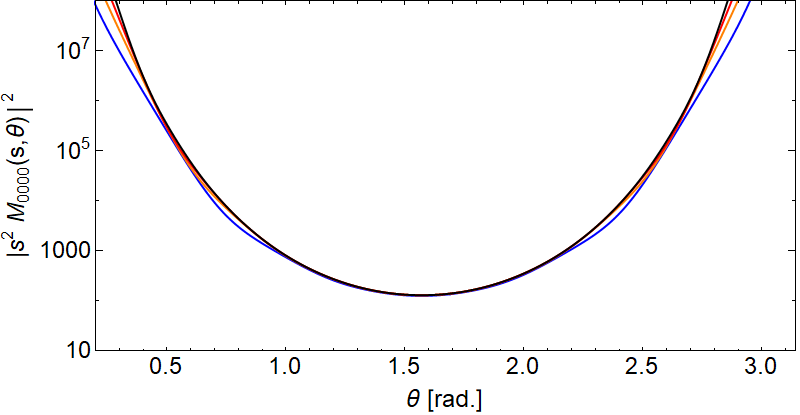}
        \caption{$|s^2{\cal M}_{0000}|^2$}
        \label{fig:7d}
    \end{subfigure}
    \begin{subfigure}{0.40\textwidth}
        \centering
        \includegraphics[width=\textwidth]{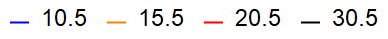}
    \end{subfigure}
    \caption{Angular dependence of the remaining pion scattering amplitudes multiplied by $s^2$ at selected values of ${\tilde \alpha}^\prime s$ (see legend).}
    \label{fig:predictions_other}
\end{figure}

We plot our predictions for the pion scattering amplitudes in figure \ref{fig:predictions_other}. The only amplitude which has a definite zero is the $\pi^+\pi^-\to\pi^+\pi^-$ amplitude. As such other amplitudes are in fact not expected to exhibit the same dips. Even without these zeros, there are other features such as local maxima and minima in the amplitudes ${\cal M}_{0000}$ and ${\cal M}_{+0+0}$ that could be seen as predictions of our model. On the other hand, these tend to be at small $t/s$ (or $u/s$), so a refinement on the model to outside the strict fixed-angle regime would be important to resolve them.

\begin{figure}[hbt!]
    \centering
    \begin{subfigure}[b]{0.45\textwidth}
        \centering
        \includegraphics[width=\textwidth]{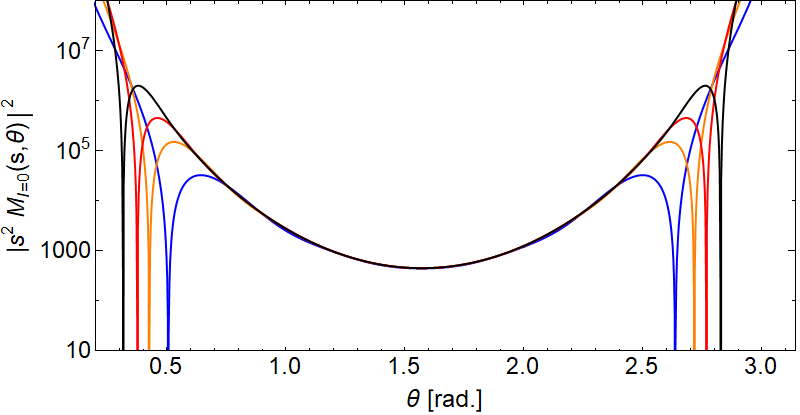}
        \caption{$s^{2} {\cal M}[I=0]$}
        \label{fig:9a}
    \end{subfigure}
    \hfill
    \begin{subfigure}[b]{0.45\textwidth}
        \centering
        \includegraphics[width=\textwidth]{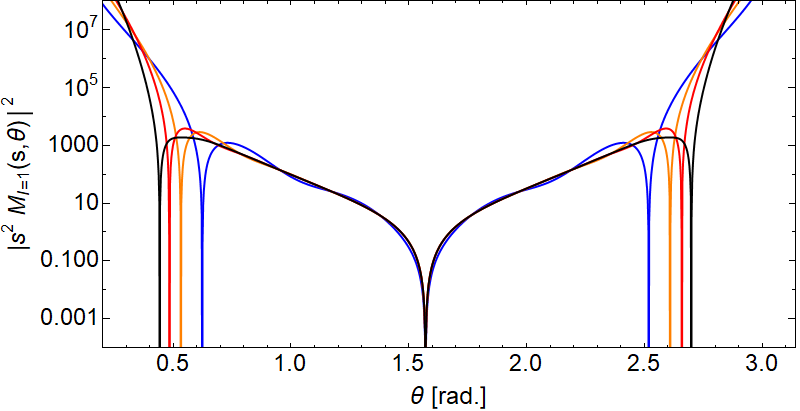}
        \caption{$s^{2} {\cal M}[I=1]$}
        \label{fig:9b}
    \end{subfigure}
     \begin{subfigure}{0.40\textwidth}
        \centering
        \includegraphics[width=\textwidth]{figures/predictions_legend.png}
    \end{subfigure}
    \caption{Pion amplitudes of channels with fixed isospin.}
    \label{fig:pion_amplitudes_isospin}
\end{figure}

In figure \ref{fig:pion_amplitudes_isospin} are plotted the amplitudes for the definite isospin channels $I = 0$ and $1$. The remaining channel $I = 2$ is the ${\cal M}_{++++}$ amplitude plotted before in figure \ref{fig:predictions_other}. 

\clearpage
\section{Summary and Outlook} \label{sec:5_conclusions}

In this paper, we obtained predictions for the angular dependence of 4-point pion scattering amplitudes following the holographic approach of reference \cite{Armoni:2024gqv}, in the specific case of the hard-wall AdS model. We compared our predictions with experimental data for $\pi^{+} \pi^{-} \to \pi^{+} \pi^{-}$ scattering.\,\,We focused primarily on the high-energy fixed-angle regime, though we also discussed the Regge regime.\,\,
Before summarising our findings, we briefly review some of the most important limitations of both our approach and the experimental data.

The ansatz for our pion scattering amplitudes, taken from reference \cite{Armoni:2024gqv}, was constructed based on the work of Polchinski and Strassler \cite{Polchinski:2001tt}.\,\,Our predictions derived from this ansatz are restricted to tree-level scattering, with all calculations carried out at leading order in the supergravity approximation $\alp / R_5^2 \ll 1$.\,\,With this approach, we successfully recovered the constituent counting rule and Regge behaviour. Since our predictions rely
solely on the asympotically AdS region of the hard-wall geometry, which models QCD’s UV-behaviour via a CFT, they are expected to be reliable only in the high-energy fixed-angle limit.\,\,In addition, note that our ansatz does not account for contributions from the WZW term (although none arise from vertices with two pions and an axial-vector meson, it would be worth checking whether vertices involving two pions and excited string states contribute), and quark masses are neglected.

On the other hand, the experimental data taken from reference \cite{Bromberg:1983he}, against which we compared our predictions, is of limited reliability for pion scattering, as it was extracted from a baryon-pion scattering experiment.\,\,Moreover, since additional phenomenological models were required to relate it to pion scattering, the validity of the extracted data beyond a qualitative level is limited.

Despite the various limitations outlined above, the qualitative agreement observed in \hyperref[sec:4_1]{Section 4.1} between our approach and the extracted data in the high-energy fixed-angle regime is encouraging. However, our approach is expected to be reliable only in the strict high-energy fixed-angle limit, whereas our comparisons at finite $s$ require fixing the parameter $\tilde{\alpha}^{\prime}$. As discussed in \hyperref[sec:4_1]{Section 4.1}, together with input from phenomenology, we fixed the value of this parameter. Given this procedure, the reader may question the reliability of our predictions. As we emphasized in \hyperref[sec:4_1]{Section 4.1}, however, fixing $\tilde{\alpha}^{\prime}$ is necessary to set the scale for the onset of the high-energy behaviour. Moreover, the fits that qualitatively match the data also qualitatively match the asymptotic behaviour exhibited by our pion scattering amplitudes in the strict high-energy fixed-angle regime. On this basis, we argue that our approach exhibits qualitative agreement with the data in the high-energy fixed-angle regime.

Our work can be extended and improved in several directions. Possible improvements include incorporating internal-space momentum contributions ({\it i.e.} the components $p_w^{(i)}$), a nontrivial dilaton profile, or quark masses. Further refinements will be especially valuable once more accurate pion-scattering data become available in the high-energy regime. Addressing some of these issues will likely require a better understanding of the Polchinski–Strassler proposal \cite{Polchinski:2001tt} in string theory.

Our ansatz also shares similarities with that of reference \cite{Veneziano:2017cks}, where a pion scattering amplitude based on infinitely many strings with varying tensions was proposed. It would be interesting to further develop this model and compare with ours and with the experimental data.

Recently, pion scattering amplitudes have also been computed using bootstrap techniques \cite{Guerrieri:2024jkn}. It would be interesting to compare these results with our approach, if there is a common range of validity.

\acknowledgments{
We are grateful to S. Sugimoto for insightful discussions and collaboration.\,\,We thank B. Pyszkowski for collaboration and discussion. We thank M. Bianchi, Z. Komargodski, T. Sakai, M. Watanabe, M. Ward, N. Zenoni, R. Wendell, S. K. Sake, J. Barman, J. Nongmaithem, and X. U. Nguyen for discussions.

DW was supported by an INFN postdoctoral fellowship and the INFN project ST\&FI “String Theory and Fundamental Interactions”.}

\newpage

\appendix
\section{Supplementary Figures} \label{app:A_graphs}
In this appendix, we include  supplementary figures to \hyperref[sec:4_comparison]{Section 4}.

\begin{figure} [h!] \centering
	\begin{subfigure}[b]{0.45\textwidth}
		\includegraphics[width=\textwidth]{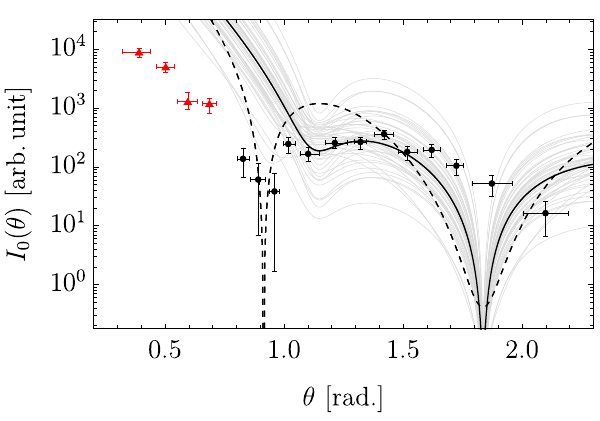}\caption{$\sqrt s = 2.1$--$2.3$ GeV} 
	\end{subfigure}
	\begin{subfigure}[b]{0.45\textwidth}
		\includegraphics[width=\textwidth]{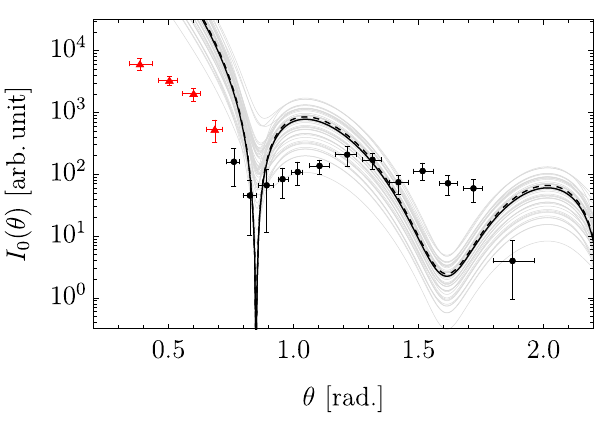}\caption{$\sqrt s = 2.3$--$2.5$ GeV}
	\end{subfigure}
	
	\vspace{0.2cm}
	
	\begin{subfigure}[b]{0.45\textwidth}
		\includegraphics[width=\textwidth]{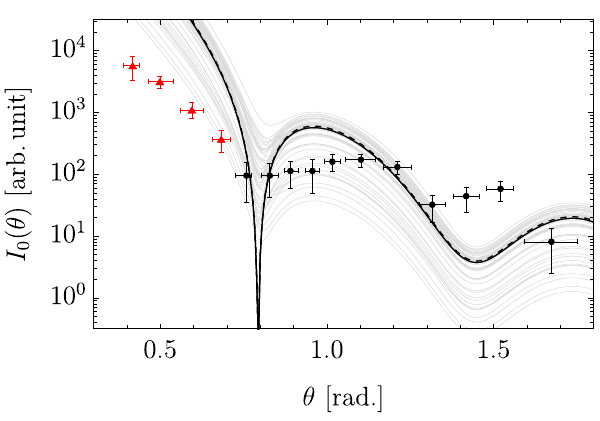}\caption{$\sqrt s = 2.5$--$2.7$ GeV} 
	\end{subfigure}
	\begin{subfigure}[b]{0.45\textwidth}
		\includegraphics[width=\textwidth]{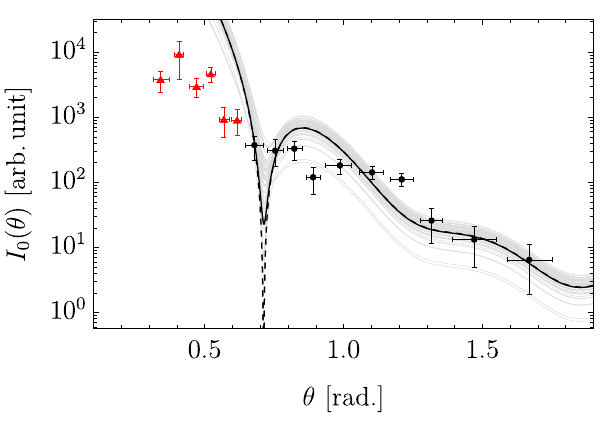}\caption{$\sqrt s = 2.7$--$3.1$ GeV}
	\end{subfigure}
	\caption{Fits for 100 GeV dataset. See Caption of \hyperref[fig:fits175]{Figure 5} for description.}
	\label{fig:fits100}
\end{figure}

In \hyperref[fig:fits100]{Figure 9} are presented the fits to the 100 GeV data set of \cite{Bromberg:1983he}. The fitting parameters are listed in \hyperref[tab:fitparams100]{Table 2} in the main text.

\section{Evidence for 4-Point Contribution in Experimental Data} \label{app:B_validation}

In this appendix, we plot some of the data for the component $I_{0}$ from the $\pi^{-} p \to \pi^{+} \pi^{-} n$ measurements reported in reference \cite{Bromberg:1983he} (for the definition of this component refer back to \hyperref[sec:3_experimental]{Section 3}). For the same binned values of $s$, we superimpose the measurements from the 100 GeV and 175 GeV data sets. At a qualitative level, the behaviour at each binned value of $s$ appears to show strong resemblance. We take this to suggest that the process $\pi^{-} p \to \pi^{+} \pi^{-} n$ includes a significant contribution from an underlying 4-point scattering mechanism. This observation was already previously noted in reference \cite{Bromberg:1983he}; however, here we take the opportunity to plot the different data sets superimposed for direct comparison.

\begin{figure} [h!] \label{fig:7}
	\centering
	\begin{subfigure}[b]{0.45\textwidth}
		\includegraphics[width=\textwidth]{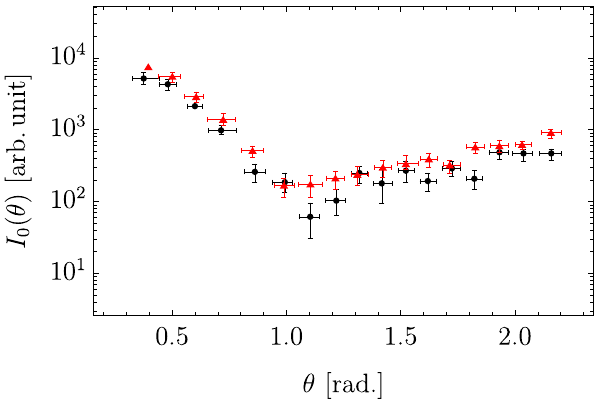}
		\caption{$\sqrt s = 1.7$--$1.9$ GeV} 
	\end{subfigure} 
	\begin{subfigure}[b]{0.45\textwidth}
		\includegraphics[width=\textwidth]{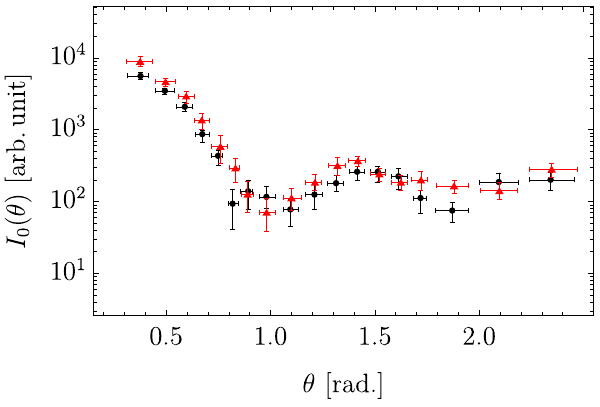}
		\caption{$\sqrt s = 1.9$--$2.1$ GeV}
	\end{subfigure}
	\vspace{0.5cm}
	
	\begin{subfigure}[b]{0.45\textwidth}
		\includegraphics[width=\textwidth]{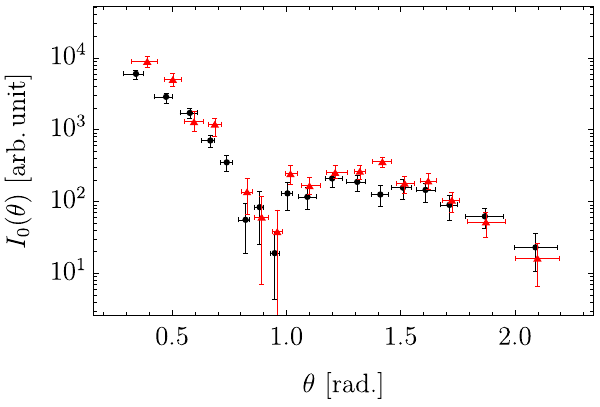} 
		\caption{$\sqrt s = 2.1$--$2.3$ GeV} 
	\end{subfigure}
	\begin{subfigure}[b]{0.45\textwidth}
		\includegraphics[width=\textwidth]{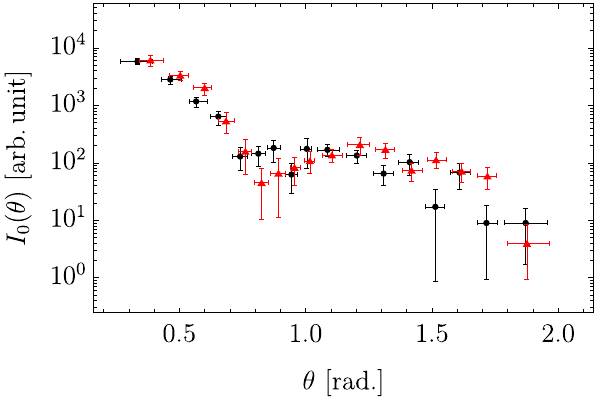}
		\caption{$\sqrt s = 2.3$--$2.5$ GeV}
	\end{subfigure}
	\vspace{0.5cm}
	
	\begin{subfigure}[b]{0.45\textwidth}
		\includegraphics[width=\textwidth]{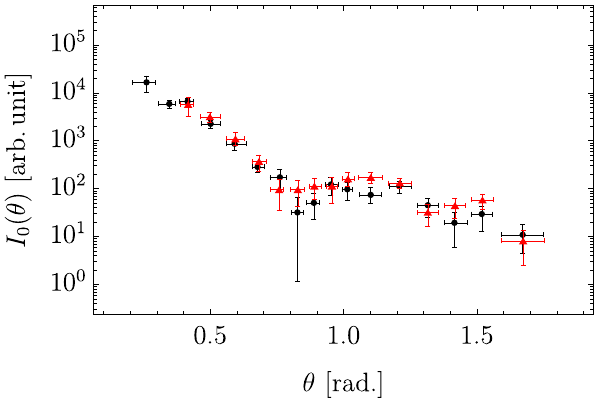}
		\caption{$\sqrt s = 2.5$--$2.7$ GeV}
	\end{subfigure}
	\caption{Comparisons of the component $I_{0}$ from the $\pi^{-} p \to \pi^{+} \pi^{-} n$ measurements reported in reference \cite{Bromberg:1983he} are shown for different binned $s$-values. The red data points correspond to the 100 GeV set, while the black data points correspond to the 175 GeV set. The data were extracted from Figure 29 in reference \cite{Bromberg:1983he}. Note, however, that we were unable to recover all of the data points, and not all error bars could be obtained with sufficient precision; such instances were, however, rare.}
\end{figure}

\bibliographystyle{JHEP}
\bibliography{biblio}

\end{document}